\documentclass{jfm}
\usepackage{graphicx} 
\usepackage{epstopdf, epsfig}
\usepackage{multirow}
\usepackage{graphicx}
\usepackage{newtxtext}
\usepackage{newtxmath}
\usepackage{natbib}
\usepackage{hyperref}
\usepackage{xcolor}

\shorttitle{High pressure bubble departure and sliding}
\shortauthor{A. Kossolapov, M. T. Hughes, B. Phillips and M. Bucci}

\title{Bubble departure and sliding in high-pressure flow boiling of water}

\author{Artyom Kossolapov\aff{1},
  Matthew T. Hughes\aff{1},
  Bren Phillips\aff{1}
 \and Matteo Bucci\aff{1}\corresp{\email{mbucci@mit.edu}}}

\affiliation{\aff{1}Department of Nuclear Science and Engineering, Massachusetts Institute of Technology, Cambridge MA 02138, USA}

\begin{document}
\maketitle

\begin{abstract}
Bubble growth, departure and sliding in low-pressure flow boiling has received considerable attention in the past. However, most applications of boiling heat transfer rely on high-pressure flow boiling, for which very little is known, as experimental data are scarce and very difficult to obtain. In this work, we conduct an experiment using high-resolution optical techniques. By combining backlit shadowgraphy and phase-detection imaging, we track bubble shape and physical footprint with high spatial (6 $\mu$m) and temporal (33 $\mu$s) resolution, as well as bubble size and position as bubbles nucleate and slide on top of the heated surface. We show that at pressures above 1 MPa bubbles retain a spherical shape throughout the growth and sliding process.  We analytically derive non-dimensional numbers to correlate bubble velocity and liquid velocity throughout the turbulent boundary layer and predict the sliding of bubbles on the surface, solely from physical properties and bubble growth rate. We also show that these non-dimensional solutions can be leveraged to formulate elementary criteria that predict the effect of pressure and flow rate on bubble departure diameter and growth time. 
\end{abstract}


\section{Introduction}
\label{sec:introduction}

Boiling is an exceptionally effective heat transfer process that can achieve some of the highest heat transfer coefficients across all heat transfer modes \citep{bergman_fundamentals_2020}. Subcooled flow boiling in particular is widely used for thermal management and energy conversion in power generation systems, such as nuclear reactors. Typically, power generation systems operate at high pressure (e.g., $\sim$10 MPa) to increase the boiling temperature of the operating fluid (e.g., water) and, consequently, the thermodynamic efficiency. Forced flow (e.g., $\sim$1000 kg m$^{-2}$ s$^{-1}$) is used to enhance boiling heat transfer and prevent boiling crises \citep{tong_prediction_1967}. However, despite decades of operational experience with high-pressure flow boiling systems, the physics of the boiling process in such systems is still poorly understood. Thus, their design still relies on empirical correlations that have a narrow range of applicability and a high degree of uncertainty. These correlations are often based on expensive experiments that duplicate the geometry, size and operating conditions of actual applications. They cannot be reliably applied to new designs without new application-specific experimental data.

By understanding the boiling phenomenon at the level of individual bubbles, it is possible to construct versatile heat flux partitioning (HFP) models that can be combined with multiphase computational fluid dynamics (CFD) tools to analyse any system configuration \citep{gilman_self-consistent_2017}. HFP models divide the total heat flux at the boiling surface into several components, each based on a specific heat transfer mechanism (e.g., evaporation, surface quenching, forced convection and sliding conduction). The equations used to calculate each heat flux component are developed based on our understanding of these physical mechanisms and depend on characteristic time and length scale parameters of the boiling process (e.g., bubble departure frequency and wait time, bubble departure diameter and nucleation site density). For example, the heat flux due to evaporation is directly related to the bubble departure volume, while the heat flux due to surface quenching is proportional to the bubble footprint area. Ultimately, each HFP term strongly depends on the growth, departure and sliding of bubbles on the heated surface. However, while bubble departure dynamics at high pressures has been a topic of interest, there is a lack of high-resolution data to support the development and validation of HFP models.

\begin{figure}
  \centerline{\includegraphics[width=0.7\textwidth]{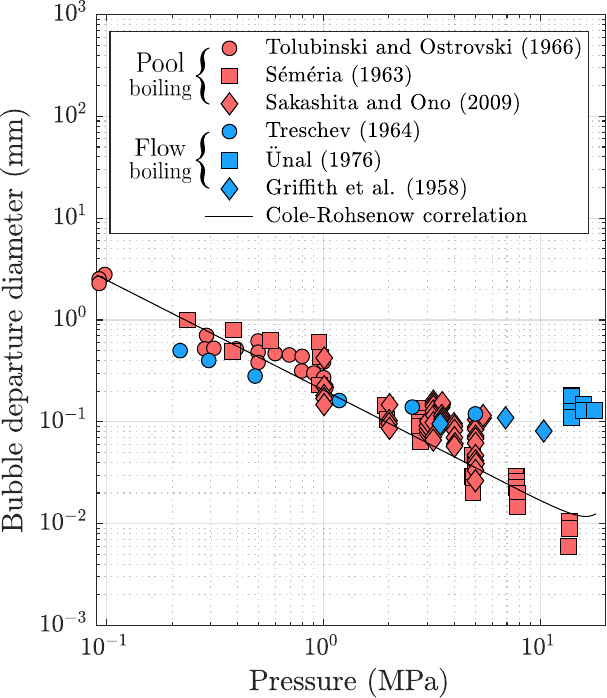}}
  \caption{Summary of high-pressure bubble departure diameter data in literature.}
\label{fig:Dd_literature}
\end{figure}

\citet{tolubinsky_mechanism_1966}, \citet{sakashita_boiling_2009} and \citet{semeria_high-speed_1963} measured bubble departure diameters in pool boiling conditions at pressures of 1.0, 5.0 and 13.7 MPa, respectively (see figure \ref{fig:Dd_literature}). In general, these data are in good agreement with semi-empirical correlations (e.g., \citet{cole_correlation_1969} obtained from \citet{kocamustafaogullari_pressure_1983}), which indicate that the departure diameter should decrease as the pressure (i.e., surface tension and liquid-vapour density ratio) decreases. However, departure diameter measurements at flow boiling conditions do not appear to follow this trend. One may observe that the departure diameter in flow boiling conditions is somehow larger than the departure diameter in pool boiling. This contradicts the idea that detaching forces induced by the forced flow (e.g., drag) should make bubbles depart faster and smaller, as shown experimentally and theoretically with refrigerants \citep{klausner_vapor_1993}. This discrepancy may arise from experimental limitations. In some cases, these measurements were obtained from still photographs that cannot be used to track the history of individual bubbles from nucleation till departure (e.g., \citet{griffith_void_1958} and \citet{treschev_number_1964}). In some other cases, they do not even represent departure diameters. For instance, \citet{unal_maximum_1976}) measured bubble diameters from a sapphire adiabatic tube downstream a 10 m long heated section, far away from the point of nucleation.

If high-pressure bubble departure diameters in flow boiling conditions were at least as small as in pool boiling conditions, the departure diameter could be 10 to 100 $\mu$m. Imaging bubbles this small requires microscopic lenses with short working distances and optical access, which is challenging to arrange in high pressure and temperature experiments. To the authors’ best knowledge, there is a severe lack of high-resolution experimental measurements of bubble parameters at the pressure and temperature of typical industrial boilers or nuclear reactors. To close this knowledge gap, we have developed an experimental test facility and optical measurement techniques to accurately capture growth, departure and sliding of bubbles in these operating conditions.

The present study also aims to determine the required complexity needed to describe this sliding process, and to predict bubble departure diameter and growth time. One notable mechanistic modelling framework is the force-balance technique, which uses conservation of momentum to obtain an equation of motion for an individual bubble. To this end, one requires accurate knowledge of all the relevant external forces that act on the bubble; otherwise, the results may yield unphysical results. This was recently demonstrated by \citet{bucci_not-so-subtle_2021}, who performed experimental and analytical investigations of forces acting on an individual bubble during pool boiling. Their findings suggest that plausible magnitudes for all the external forces and resulting bubble acceleration can only be precisely quantified if the bubble shape is accurately known at all times. Several studies demonstrated that these conclusions also applies to quasi-static injection of bubbles through orifices (\citet{duhar_dynamics_2006}, \citet{lebon_dynamics_2018} and \citet{marco_experimental_2015}). One major source of uncertainty comes from contact line surface tension forces. At sufficiently high Bond, Weber, or Capillary numbers, the bubble shape may become asymmetric. In this case, the contact line surface tension force also becomes asymmetric and prevents the bubble from sliding. The difficulty with quantifying this force arises from its dependence on the advancing and receding contact angles ($\theta_{a}$ and $\theta_{r}$, respectively), which themselves are strong functions of the temperature and operating conditions, as well as surface composition and texture. This makes calculating asymmetric contact line surface tension forces a very difficult task that can significantly impact predictions. In fact, \citet{favre_updated_2023} showed that a small change ($\sim$5°) in the contact angle and contact angle hysteresis can impact the force balance prediction accuracy by several fold. While this striking result casts doubt on using force balance models, such issues may be alleviated in our high-pressure flow boiling conditions. Smaller bubbles (i.e., with much lower Bond, Weber and Capillary numbers) may be spherical and may depart by sliding over the heated surface (i.e., the contact line surface tension becomes negligible). We show that, under these conditions, it is possible to simplify the bubble equation of motion eliminating the need for solving complex differential equations, which, in a CFD modeling framework, would consume computational time, be difficult to implement, and might make the simulation unstable. Precisely, we have made physical considerations and order-of-magnitude analyses to obtain closed-form analytical solutions to model bubble sliding, and, importantly, to predict the measured bubble departure diameter and growth time.

\section{Experimental approach}\label{sec:experimental_approach}

\begin{figure}
  \centerline{\includegraphics[width=\textwidth]{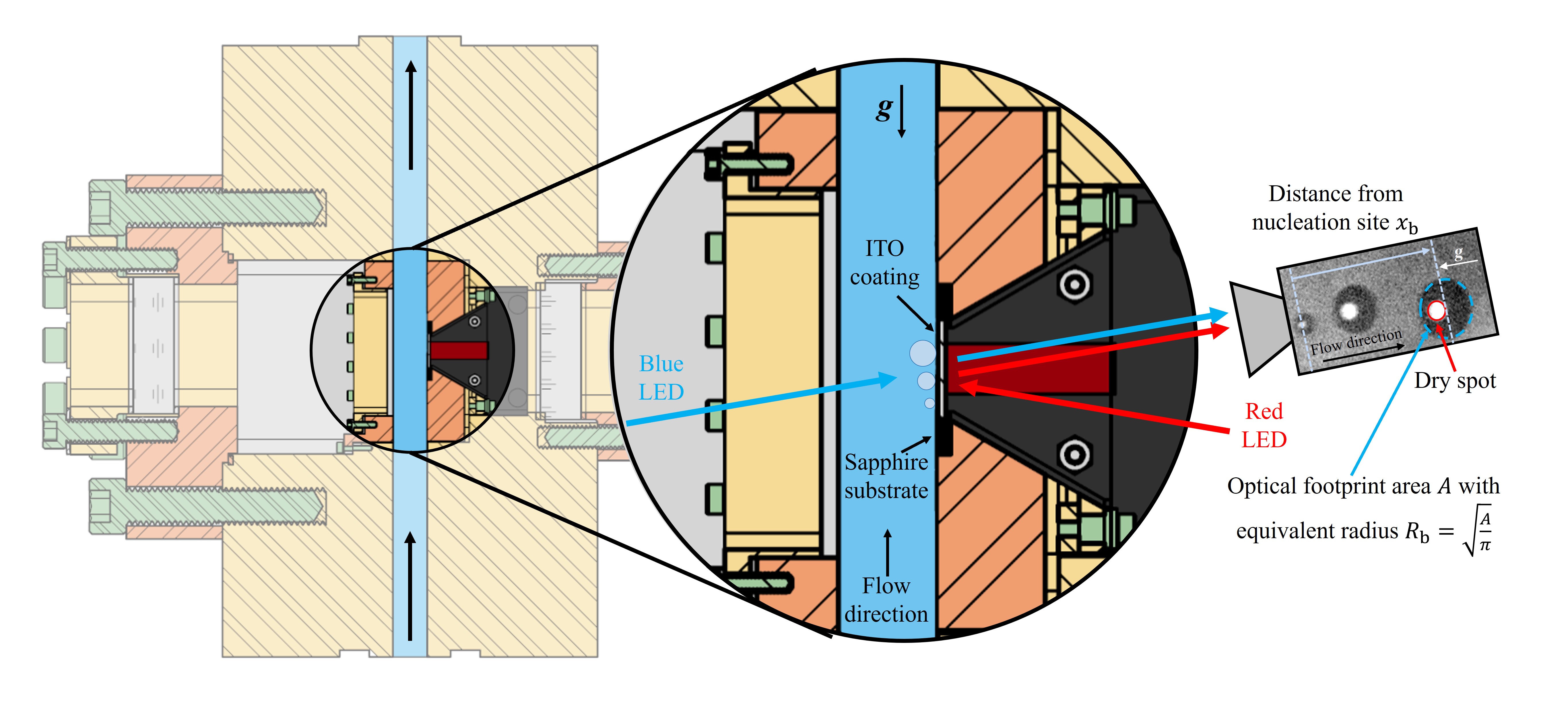}}
  \caption{Longitudinal cross-section view of test section and zoomed in view near boiling surface to illustrate optical measurement technique.}
\label{fig:test_section}
\end{figure}

We run our experiments using the test section shown in figure \ref{fig:test_section}. A detailed description of the facility can be found in \citet{kossolapov_experimental_2021}. A short description is provided hereafter. More details on the flow loop capabilities and operation can be found in the appendix. 

Briefly, the test section features a square channel with a hydraulic diameter $D_\textrm{h}$ of 11.78 mm. It is forerun by a straight vertical entrance channel with the same cross-sectional area. This entrance channel is 765 mm long (i.e., 65 hydraulic diameters) to ensure that the flow is hydrodynamically developed at the test section inlet. In our operating conditions (see table \ref{tab:test_conditions}), the channel Reynolds number ($Re_\textrm{ch}=GD_\textrm{h}/\mu_\textrm{l}$) ranges from 36800 to 157000, implying that the flow is fully turbulent \citep{White2008}. 

Each side of the test section channel has optical access, provided by sapphire windows. One of the sides features a sapphire window coated with a thin, electrically conductive indium tin oxide (ITO) layer. This ITO layer is in contact with the flow, and it is used to release, by Joule effect, the heat to boil water. The ITO electric power, and consequently the boiling heat flux, is adjusted to control the number of active nucleation sites on the ITO surface. In each test (i.e., at each operating condition), the heat flux is adjusted to only produce non-interacting, discrete bubbles that we can track using our optical technique.

The optical technique used in this study uses a combination of phase-detection and backlit shadowgraphy \citep{kossolapov_can_2021}. A blue LED is used to back lit the boiling surface. The light easily passes through the liquid but is blocked by bubbles, casting a shadow that can be easily tracked by a high-speed video (HSV) camera (Phantom v2512 recording up to 30,000 frames per second with a pixel resolution of 6 $\mu$m) positioned behind the heater (as sketched in figure \ref{fig:test_section}). At the same time, a red LED lights the boiling surface from the same side as the camera. Where the surface is in contact with liquid, the red LED light is mostly transmitted. Only a small fraction of the incident beam is reflected back and captured by the high-speed video camera. This happens because the index of refraction of sapphire and liquid is similar. Conversely, the vapour has a much different refraction index. Thus, where the surface is covered by vapour, the light is almost entirely reflected and captured by the high-speed video camera. Briefly, dry patches that form at the base of the bubble appear as bright spots in the high-speed video images. The presence of a microlayer (i.e., a very thin liquid layer that may form when a bubble grows on top of the heated surface), if any, would create multiple reflection generating interference fringes in the images. However, we did not detect any microlayer in these high-pressure flow boiling experiments.

A video is recorded in steady-state conditions for each combination of pressure and mass flux listed in table \ref{tab:test_conditions}. After the test, HSV images are processed to track the history of individual bubbles generated from specific nucleation sites. By post-processing these images, we can measure the physical size of bubbles, represented by the equivalent circular bubble radius of the optical footprint, as sketched in figure \ref{fig:test_section}. By tracking bubbles over time, we can also measure their displacement from the nucleation site until they either leave the field of view. Details on the bubble size measurement and tracking algorithm can be found in the appendix. The number of bubble histories tracked for each operating condition can be found in table \ref{tab:test_conditions}. 

\begin{table}
  \begin{center}
\def~{\hphantom{0}}
  \begin{tabular}{cccccc}
      Nominal pressure  & Mass flux $G$    &   
      Heat flux & $Re_\textrm{ch}=\frac{GD_\textrm{h}}{\mu_\textrm{l}}$  & 
      Subcooling & Number of bubbles tracked\\
      (MPa)  &(kg m$^{-2}$ s$^{-1}$)    &(kW m$^{-2}$) &   &(K)    & \\[6pt]
       \multirow{3}{*}{1}   & 500  & 514 & 36800  & 12.6  & 171\\
			     & 1000 & 503 & 73600   & 11.6  & 250\\
			     & 2000 & 1006 & 147300  & 12.1  & 3602\\[6pt]
	\multirow{3}{*}{2}   & 500  & 178 & 44200   & 10.2  & 964\\
			     & 994 & 495 & 87900   & 10.3  & 962\\
			     & 1504 & 489 & 132900  & 9.9  & 1122\\[6pt]
	\multirow{3}{*}{4}   & 500  & 291 & 52900   & 10.3  & 803\\
			     & 988 & 361 & 104600   & 10.2  & 1943\\
			     & 1500 & 613 & 158900  & 10.2  & 2424\\[6pt]
  \end{tabular}
  \caption{Operating conditions under investigation and number of individual bubble histories tracked.}
  \label{tab:test_conditions}
  \end{center}
\end{table}

\section{Results and discussion}\label{sec:results_discussion}
\subsection {Bubble appearance and growth}
\begin{figure}
  \centerline{\includegraphics[width=\textwidth]{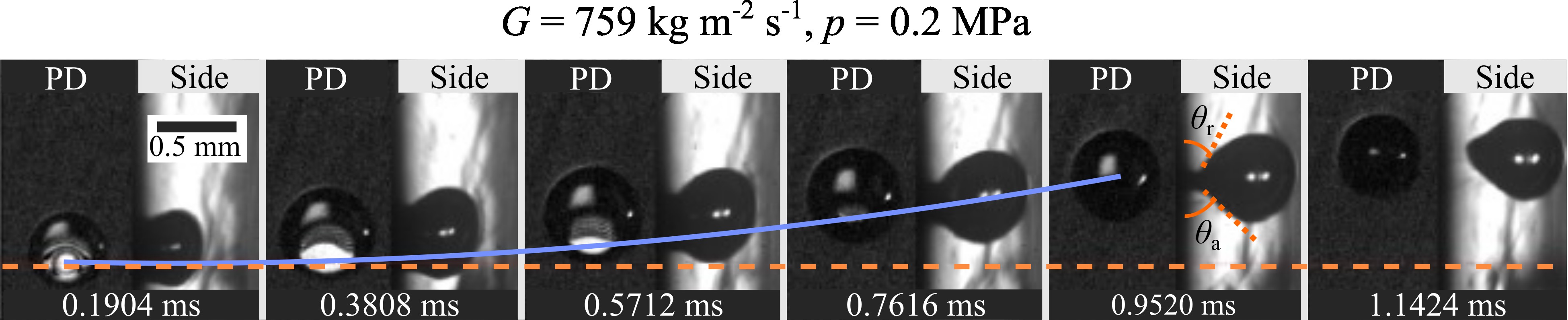}}
  \caption{Phase-detection (PD) and shadowgraphy side view images of flow boiling at 0.2 MPa. The dashed orange line represents the location of a nucleation site. The solid blue line shows the vertical displacement of a bubble over time.}
\label{fig:PD_lowp}
\end{figure}

\begin{figure}
  \centerline{\includegraphics[width=\textwidth]{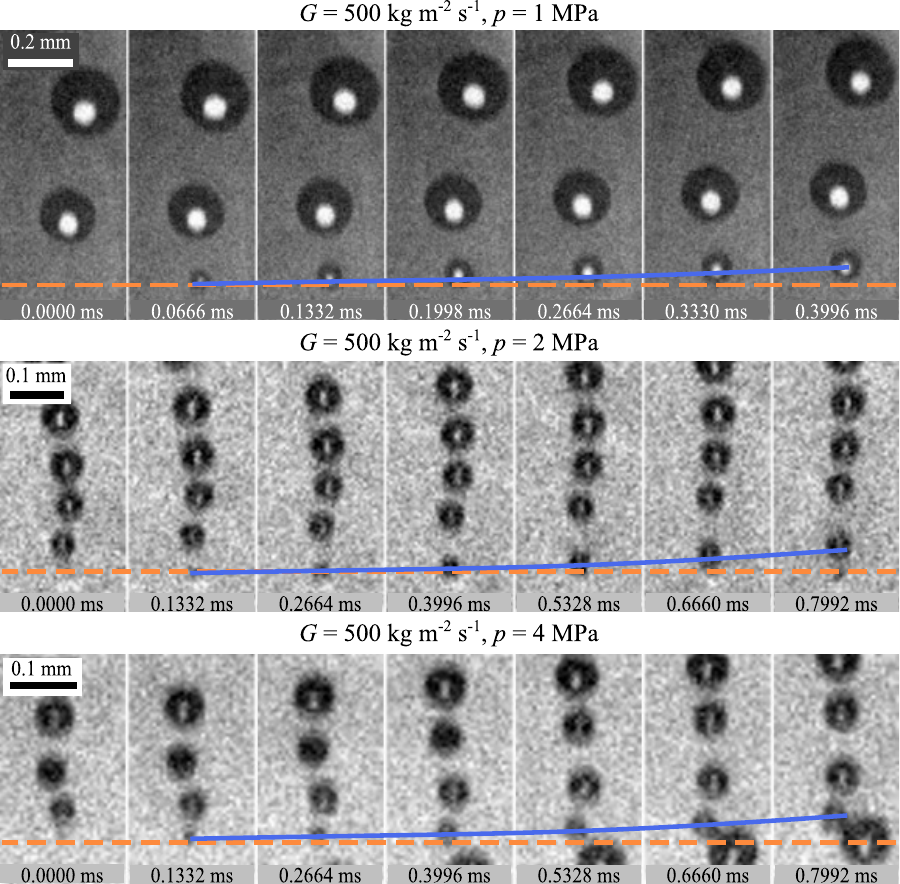}}
  \caption{ PD images of flow boiling at different pressures. The dashed orange line represents the location of a nucleation site. The solid blue line shows the vertical displacement of a bubble over time.}
\label{fig:PD_highp}
\end{figure}

Figure \ref{fig:PD_lowp} shows a series of phase-detection (labelled PD) images during flow boiling at 0.2 MPa, a relatively low saturation pressure, obtained from the study by \citet{kossolapov_experimental_2021}. As revealed by these images, and also shown by \citet{sinha_microlayer_2022}, when a vapour bubble grows in low pressure conditions, it is dragged and rolled in the direction of the flow, creating an asymmetric microlayer (shown as white fringes) and dry spot (shown in solid white). This asymmetric shape leads to contact line surface tension forces that inhibits bubble sliding. This bubble asymmetry can also be seen in the side view images, where there is a clear difference in the advancing and receding contact angle. These observations are in striking contrast to the images of vapour bubbles at high pressures collected in this study and shown in figure \ref{fig:PD_highp}. At high pressures, we did not observe a microlayer, the optical footprint is rather circular, and the physical footprint appears to be close to the centre of the bubble, indicating spherical symmetry.  Even at the lowest pressure (i.e., 1.0 MPa), the physical footprint is located near the centroid of the optical footprint up until the bubble leaves the nucleation site. The physical size of the bubbles at the moment of departure from the nucleation site also appears to be much different. For instance, the equivalent departure radius of vapour bubbles at 0.2 MPa is approximately 0.25 mm, while the departure radii of bubbles at high pressures are well below 0.05 mm. The assumption of spherical symmetry can be validated by comparing the magnitude of surface tension and other forces (i.e., buoyancy, viscous and inertial forces) acting on the bubble as pressure increases. Consider the Bond, Capillary and Weber numbers given by:
\begin{equation}
  Bo = \frac{(\rho_{\text{l}}-\rho_{\text{v}})gD_{\text{d}}^{2}}{\sigma},
  \label{Bond}
\end{equation}
\begin{equation}
  Ca = \frac{\mu_{\text{l}} U_{\text{l}}}{\sigma},
  \label{Capillary}
\end{equation}
and
\begin{equation}
  We = \frac{\rho_{\text{l}} U_{\text{l}}^{2}D_{\text{d}}}{\sigma},
  \label{Weber}
\end{equation}
respectively, where the departure diameter $D_{\text{d}}$ is the characteristic length and the local liquid velocity at the bubble centroid $U_{\text{l}}$ is the characteristic velocity. From 0.2 MPa to 1.0 MPa, for example, surface tension decreases by 30\%, liquid density by 6\%, liquid viscosity by 60\%, while vapour density increases by 78\%. Ultimately, the most important effect is the decrease in the bubble departure diameter, which drops by a factor of five, resulting in the Bond number decreasing by a factor of 20. With pressure, the characteristic velocity $U_{\text{l}}$ will drop according to the bubble diameter; therefore, the Capillary number and Weber number will also decrease. Other factors that may cause spherical asymmetry, such as shear \citep{Taylor1934} or turbulence deformations \citep{Ni2024}, should also be negligible in our conditions. Overall, dimensional arguments and experimental observations both indicate that bubbles should become more spherical as pressure increases. Further discussion and more quantitative justification of these assumptions can be found in the appendix.

\begin{figure}
  \centerline{\includegraphics[width=\textwidth]{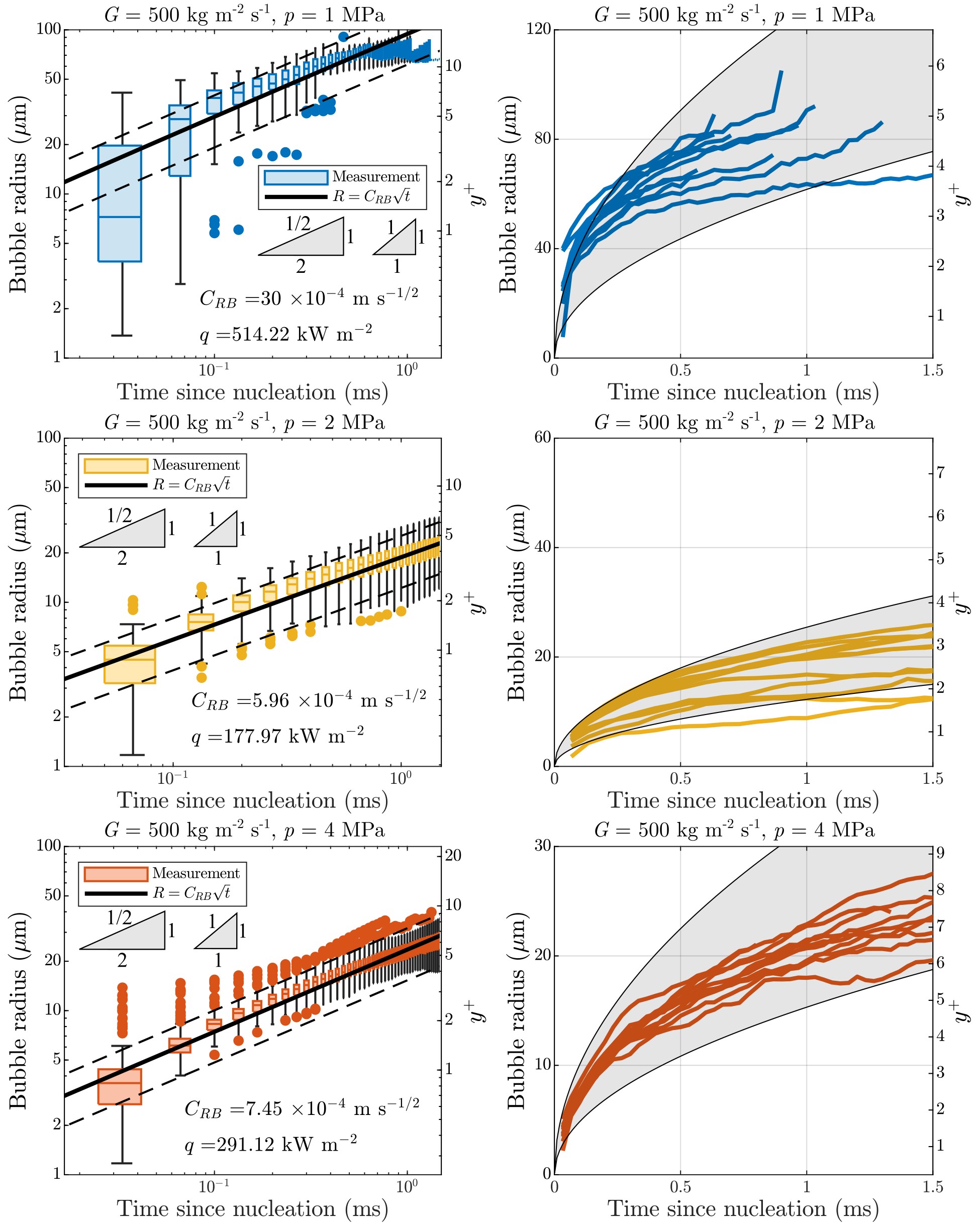}}
  \caption{Bubble growth statistics (left) and selected histories (right) from 1 to 4 MPa at a mass flux of 500 kg m$^{-2}$ s$^{-1}$.}
\label{fig:growth_lowG}
\end{figure}

\begin{figure}
  \centerline{\includegraphics[width=\textwidth]{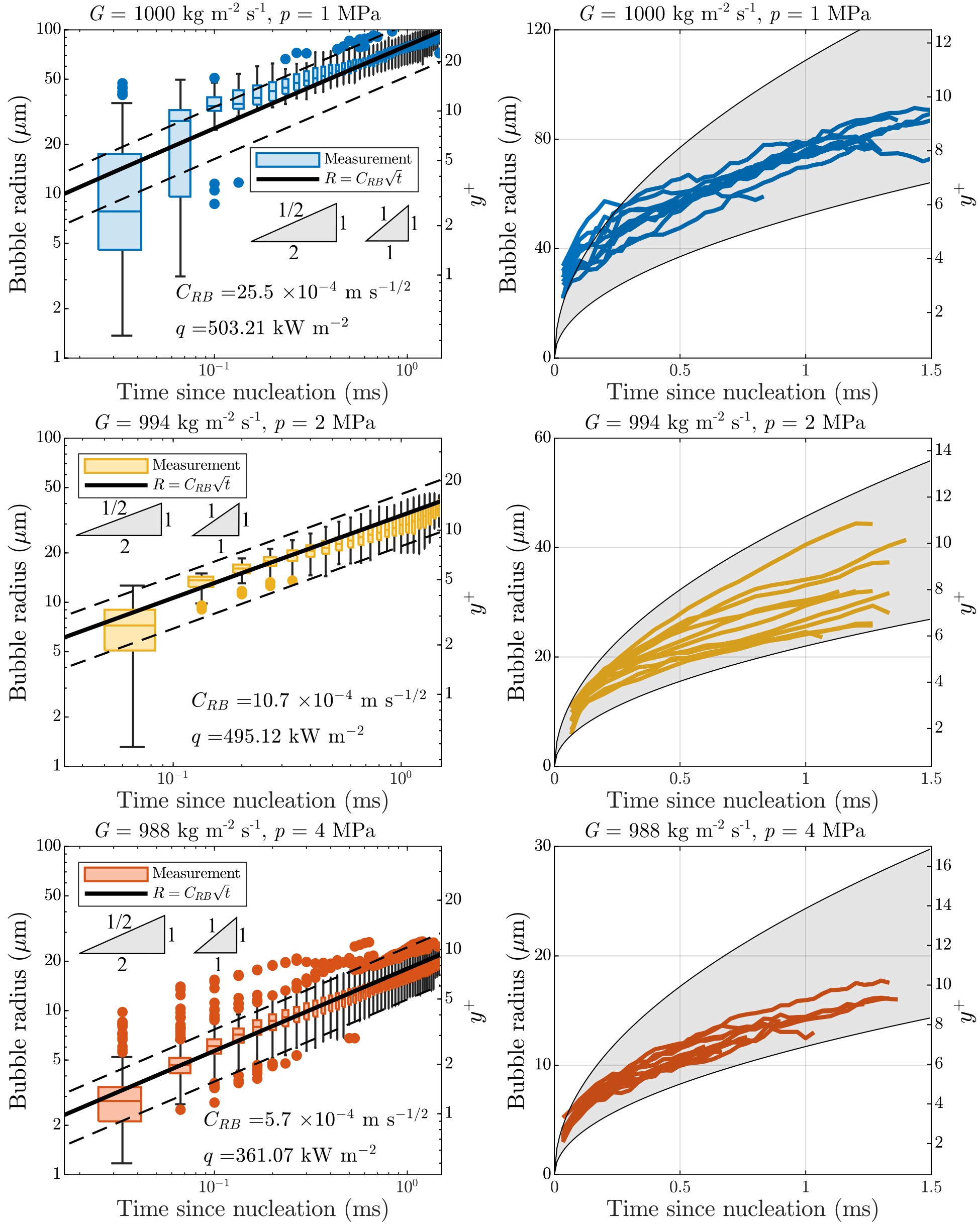}}
  \caption{Bubble growth statistics (left) and selected histories (right) from 1 to 4 MPa at a mass flux of 1000 kg m$^{-2}$ s$^{-1}$.}
\label{fig:growth_medG}
\end{figure}

\begin{figure}
  \centerline{\includegraphics[width=\textwidth]{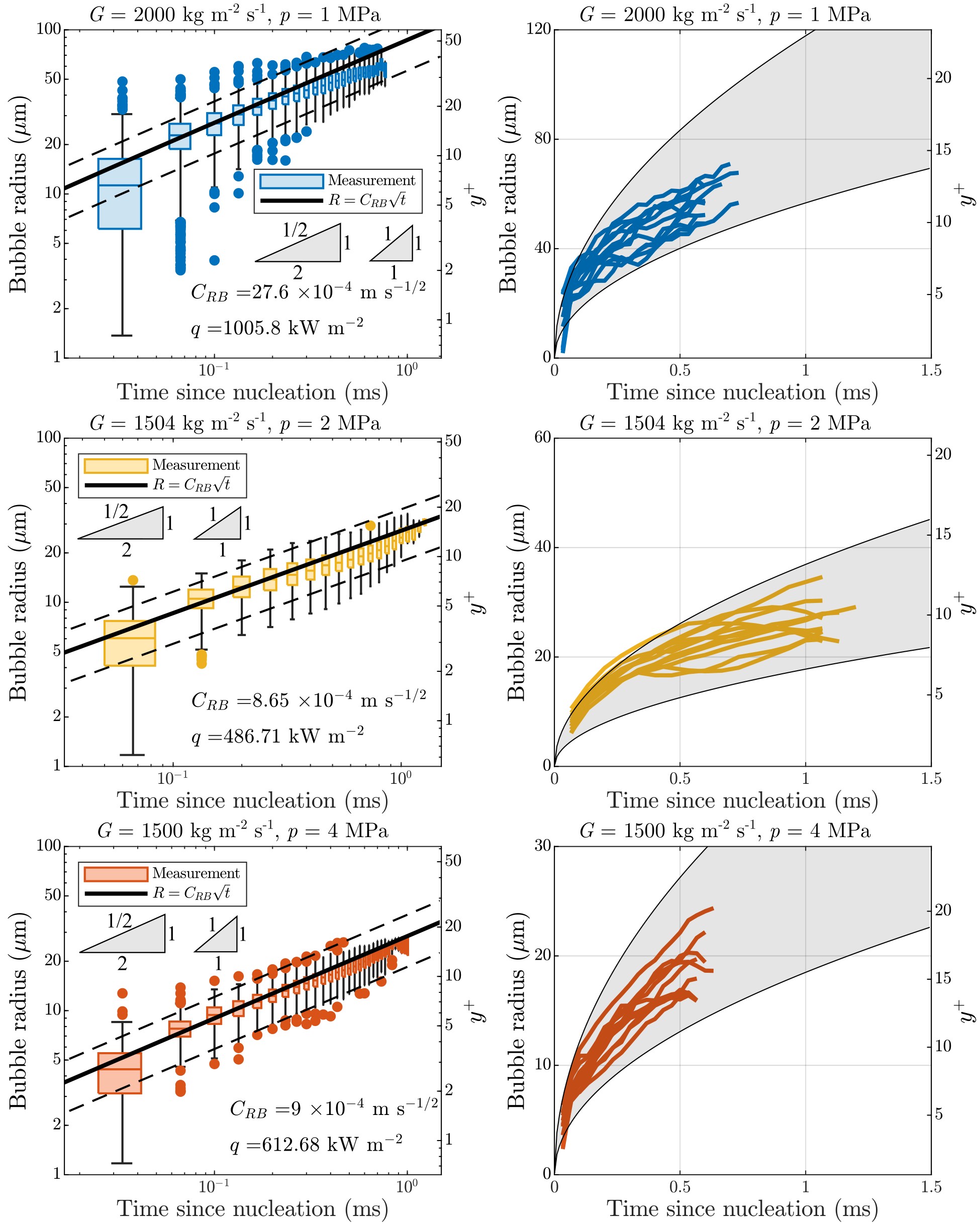}}
  \caption{Bubble growth statistics (left) and selected histories (right) from 1 to 4 MPa at the highest mass fluxes tested (2000 kg m$^{-2}$ s$^{-1}$ at 1 MPa and 1500 kg m$^{-2}$ s$^{-1}$ at 2-4 MPa).}
\label{fig:growth_highG}
\end{figure}

To model bubble departure and sliding, we must account for the change in bubble size with time. Therefore, we measure the bubble radius over time as it grows and slides on the surface. Figures \ref{fig:growth_lowG}-\ref{fig:growth_highG} summarizes the statistical distribution of bubble sizes with time for each operating pressure, colored as blue, yellow and red for 1, 2 and 4 MPa, respectively, where the last bubble size measurement occurs when it leaves the high-speed camera field of view. Data are grouped by mass flux: 500 kg m$^{-2}$ s$^{-1}$ in figure \ref{fig:growth_lowG}, 1000 kg m$^{-2}$ s$^{-1}$ in figure \ref{fig:growth_medG} and 1500 or 2000 kg m$^{-2}$ s$^{-1}$ in figure \ref{fig:growth_highG}. At each frame, or time step, the equivalent bubble radius is visualized using a box and whisker plot, alongside dots to show any statistical outliers. To the right of the statistical distributions are samples of individual bubble growth histories, depicted using the same color scheme noted above.

Unfortunately, due to the of lack wall and local fluid temperature measurements, we cannot formulate a model of the bubble growth. Instead, we provide a simple but physically motivated and tractable semi-empirical correlation that will be used as an input in our mechanistic departure and sliding models. The measurements show that the equivalent bubble radius is mostly proportional to the square-root of time, and can be fitted to the equation:
\begin{equation}
  R_{\text{b}} = C_{\text{RB}}\sqrt{t}
  \label{R_CRB}
\end{equation}
where $C_{\text{RB}}$ is an empirical constant.  The value of $C_{\text{RB}}$ is chosen so that the mean absolute error (MAE) between equation \ref{R_CRB} and instantaneous bubble radius distributions, shown as box and whisker plots in figures \ref{fig:growth_lowG}-\ref{fig:growth_highG}, is minimized. Equation \ref{R_CRB} is shown as a solid black line over the experimental data in the left plots of figure 5. Dashed black lines corresponding to an uncertainty in $C_{\text{RB}}$ equal to ±35\% are also shown for illustrative purposes. We note that it could be possible to correlate bubble growth with other power-laws (e.g., $R_\textrm{b} = a t^{b}$), but there would be no major improvement in the prediction accuracy compared to the square-root of time fit. Instead, a square-root of time dependency is physically motivated and hints toward a heat-diffusion controlled growth. As discussed by \citet{mikic_bubble_1970}, bubble growth can be limited by two mechanisms, namely inertia and heat diffusion from the superheated liquid. Since the timescale associated with inertial-controlled bubble growth is on the order of nanoseconds for our condition, we expect bubble growth to be heat-diffusion controlled, which should result in the bubble radius increasing with the square-root of time. Other effects, such as the translational motion of the bubble, can increase the growth rate \citep{Legendre1998Thermal}, but such effects are only pronounced at high $\Rey_\text{b}$, which is not the case in our conditions. With these considerations in mind, we refer to the work of \citet{plesset_growth_1954} to provide an interpretation of the fitting coefficient $C_\textrm{RB}$. According to \citet{plesset_growth_1954}, the heat-diffusion controlled growth of the bubble radius in a uniformly superheated liquid can be predicted by:
\begin{equation}
  R_{\text{b}} \approx B \sqrt{t} = \sqrt{\frac{12\alpha_{\text{l}}}{\pi}} \frac{\rho_{\text{l}}}{\rho_{\text{v}}}\frac{c_{\text{p,l}}}{h_{\text{lv}}} (T_{\text{w}} - T_{\text{sat}})\sqrt{t}.
  \label{R_B}
\end{equation}
Note that our $C_{\text{RB}}$ is equivalent to the theoretical parameter $B$ in equation \ref{R_B}. Since we could not measure the nucleation temperature $T_{\text{w}}$, it is impossible to evaluate the parameter $B$ experimentally. However, we can estimate $(T_{\text{w}} - T_{\text{sat}})$ based on the correlation by \citet{jens_analysis_1951} and the measured wall heat flux. The empirical $C_{\text{RB}}$, bubble radius MAE and theoretical $B$ are reported in table \ref{tab:C_RB_values}, where on average the ratio $C_{\text{RB}}$/$B$ is 0.55 ± 0.21. Our bubbles grow slower than what is predicted by equation \ref{R_B}. This is expected, as equation \ref{R_B} was obtained for a spherical bubble growing in a uniformly superheated liquid, while our bubbles grow in presence of subcooled liquid, which slows the growth down.

\begin{table}
  \begin{center}
\def~{\hphantom{0}}
  \begin{tabular}{ccccc}
      Nominal pressure  & Mass flux $G$    &   
      Empirical $C_{\text{RB}}$  & Theoretical $B$ & $C_{\text{RB}}/B$\\
      (MPa)  &(kg m$^{-2}$ s$^{-1}$)    &(10$^{-4}$ m s$^{-1/2}$)   & (10$^{-4}$ m s$^{-1/2}$) & \\[6pt]
       \multirow{3}{*}{1}   & 500  & 30.0      & 52.1  & 0.58\\
                             & 1000 & 25.5      & 52.9  & 0.48\\
                             & 2000 & 27.6      & 61.9  & 0.45\\[6pt]
        \multirow{3}{*}{2}  & 500  & 5.96      & 19.4  & 0.31\\
                             & 994  & 10.7      & 24.9  & 0.43\\
                             & 1504 & 8.65      & 25.1  & 0.34\\[6pt]
        \multirow{3}{*}{4}  & 500  & 7.45      & 8.54  & 0.87\\
                             & 988  & 5.70      & 9.01  & 0.63\\
                             & 1500 & 9.00      & 10.2  & 0.88\\[6pt]
  \end{tabular}
  \caption{Summary of empirical $C_{\text{RB}}$ measurements and its comparison against theoretical predictions.}
  \label{tab:C_RB_values}
  \end{center}
\end{table}

It can be seen both in figures \ref{fig:growth_lowG}-\ref{fig:growth_highG} that the heat-diffusion controlled growth fitting (equation \ref{R_CRB}) is less accurate at 1 MPa. This error is typically largest long after the bubble nucleated, i.e., near the end of when bubbles are tracked. This is because the bubble growth rate at 1 MPa starts to decrease and eventually appears to stop growing. Considering the fact that bubbles at 1 MPa are significantly larger than bubbles at 2 and 4 MPa, it is plausible that their size makes effects of subcooling more substantial, preventing the bubbles from growing further.  The growth statistics for 2 and 4 MPa, on the other hand, generally demonstrate better agreement, but contain more outliers. This is primarily because many more bubbles can be tracked at higher pressures due to their smaller size and shorter time near the nucleation site; therefore, more outliers will inevitably be measured. However, the relative number of outliers is small as illustrated in the sample of the red growth histories in figures \ref{fig:growth_lowG}-\ref{fig:growth_highG}, where many of the individual bubbles follow the expected growth trend quite well. On average, equation \ref{R_CRB} fits the bubble size measurements within 27\% across all operating conditions and overall, measurements indicate that bubble growth is consistently proportional to the square-root of time (i.e., heat-diffusion controlled), particularly during the times when the bubble is still near the nucleation site. However, while we acknowledge that our treatment of bubble growth is not complete, and a more precise and mechanistic growth model will be an important goal of future studies, equation \ref{R_CRB} provides a reasonable boundary condition for the sliding and departure models, which is the focus of this work.

In passing, we also observe that, for a given mass flux, an increase of pressure decreases the bubble growth rate. For instance, when increasing the pressure from 1 to 2 MPa, the liquid-vapour density ratio decreases by a factor of two, while the thermal diffusivity and Jakob number $Ja=c_{\text{p,l}} (T_{\text{w}} - T_{\text{sat}})/h_{\text{lv}}$   only decrease by 3\% and 6\% percent, respectively. The Jakob number does not change as much because the reduction of latent heat is nearly balanced out by the reduction in nucleation temperature due to decreased surface tension. Therefore, the primary reason why $C_{\text{RB}}$ decreases with increasing pressure is because of the increase in vapour density, which explains the slower bubble expansion as liquid evaporates.

Mass flux, on the other hand, does not have an obvious effect on the bubble size, which supports the numerical findings of \citet{Legendre1998Thermal}, but it of course affects the local liquid velocity near the bubble. The secondary axes of the plots in figures \ref{fig:growth_lowG}-\ref{fig:growth_highG} show the equivalent bubble radius in wall units, where the friction velocity is estimated using the friction factor correlation given by McAdams \citep{todreas_nuclear_2011} (see appendix for details). Note that, as mass flux increases, the thickness of the viscous sublayer decreases, the local fluid velocity near the bubble increases and creates stronger detaching forces that promote bubble sliding.

\subsection{Bubble sliding}
\begin{figure}
  \centerline{\includegraphics{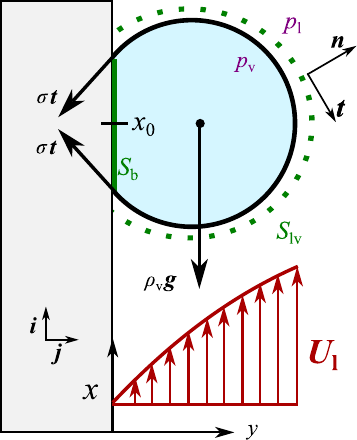}}
  \caption{Schematic of a force balance on a single, sliding vapour bubble.}
\label{fig:force_balance}
\end{figure}

We now revisit figures \ref{fig:PD_lowp} and \ref{fig:PD_highp} to discuss the observed sliding phenomenon in detail. It can be seen that the bubble continues to adhere to the boiling surface as it slides vertically upward in the direction of the bulk flow. This is evidenced by presence of a bright patch, indicating a dry spot, in the centre of the bubble optical footprint. At 0.2 MPa, bubbles do not physically lift off from the boiling surface until they are far away (0.75 mm, $\sim$1.5 $D_{\text{d}}$) from their original nucleation site \citep{kossolapov_experimental_2021}. At higher operating pressures, the distance at which bubbles lift off from the boiling surface is even larger (relative to the bubble size). This is expected because the bubble asymmetry at 0.2 MPa should yield contact line surface tension forces that reduce bubble acceleration in the vertical direction. One can also see an increase in the number of visible bubbles at a given mass flux as pressure increases from 1 to 2-4 MPa. This behavior can be explained qualitatively, as increasing pressure decreases surface tension, which in turn decreases the nucleation temperature and wait time.  
The vertical displacement of individual bubbles is illustrated with solid blue lines in figure \ref{fig:PD_highp}. It can be seen that the bubble trajectory is non-linear with time, as the bubble accelerates as it grows. One explanation for this observation is that a larger bubble is exposed to larger local fluid velocities, which aids in accelerating the bubble away from its original nucleation site position. To better understand sliding from a mechanistic point of view, a momentum balance is written for a growing vapour bubble that is surrounded by a forced flow of liquid, shown schematically in figure 6. By assuming axis-symmetric growth of a spherical bubble with negligible momentum transfer due to evaporation, the momentum balance can be written as \citep{bucci2020theoretical}:
\begin{equation}
  \rho_{\text{v}}V_{\text{b}}\frac{d\boldsymbol{U_\text{b}}}{dt}=\oiiint\limits_{V_{\text{b}}}{\rho_{\text{v}} \boldsymbol{g}} dV + \oint\limits_{CL}\sigma\boldsymbol{t}d\ell - \iint\limits_{S_{\text{b}}}p_{\text{v}}\boldsymbol{n}dS -\iint\limits_{S_{\text{lv}}}p_{\text{l}}\boldsymbol{n}dS + \iint\limits_{S_{\text{lv}}}\boldsymbol{\tau\cdot n}dS
  \label{vector_momentum_balance}
\end{equation}
where $V_{\text{b}}$ is the volume bubble, $\boldsymbol{U_\text{b}}$ is the bubble velocity, $CL$ is the contact line, and the meaning of the other symbols is obvious from figure \ref{fig:force_balance}.We assumed that the only body force is the weight of the bubble, and the only major surfaces forces are the contact line surface tension forces, viscous stresses along the liquid-vapour interface and pressure forces from the liquid and vapour phase. Because the average flow-field and gravity are in the vertical direction, only the vertical-direction of the momentum balance is needed to predict sliding. This simplification coupled with some common algebraic manipulations yields:
\begin{equation}
  \rho_{\text{v}}V_{\text{b}}\frac{dU_\text{b}}{dt}=\oint\limits_{CL}\sigma\boldsymbol{t}d\ell\boldsymbol{\cdot i} + (\rho_{\text{l}}-\rho_{\text{v}})gV_{\text{b}} -\iint\limits_{S_{\text{lv}}}(p_{\text{h}}-p_{\text{c}})\boldsymbol{n}dS\boldsymbol{\cdot i} + \iint\limits_{S_{\text{lv}}}\boldsymbol{\tau\cdot n}dS\boldsymbol{\cdot i}
  \label{x_momentum_balance}
\end{equation}
where the forces (in the right-hand side) from left to right are the contact line surface tension force, buoyancy force and total hydrodynamic force (given by the third and fourth terms). The quantity $p_{\text{h}}$ is the hydrodynamic pressure of the liquid, and  $p_{\text{c}}$ is a reference pressure, taken to be the liquid pressure at the bubble base. This is done to algebraically shift the effect of hydrostatics from the liquid pressure surface integral to the buoyancy force term (given by the second term in equation \ref{x_momentum_balance}) \citep{bucci2020theoretical}. Because most of the bubbles that are observed during sliding appear to be spherical and symmetric, the net contact line surface tension force can be neglected. The total hydrodynamic force (i.e., the last two terms on the right-hand side of equation \ref{x_momentum_balance}) is modeled as the combination of a quasi-steady drag force and a virtual (or added) mass force \citep{favre_updated_2023}. Quasi-steady drag of a bubble inside a turbulent shear flow near a wall near our conditions has yet to be measured or simulated. However, drag models for rigid spheres and clean bubbles within linear shear flows bounded by a wall (i.e., without turbulence) have been extensively studied and match our operating conditions well because most bubbles remain in the viscous sublayer (see figure \ref{fig:bubble_cdfs} (left) in the appendix). We also note that turbulence typically does not have a systematic effect on mean drag forces \citep{Wu1994, Sridhar1995, Bagchi2003, Salibindla2020}, while the presence of a wall does \citep{zeng_forces_2009, Shi2020, shi_drag_2021}; therefore, a model based on measurements or simulations of bubble drag within a linear shear flow bounded by a wall should sufficiently capture drag on the bubbles we observe. Additional effects that may impact drag, such as bubble interactions and wake effects, should also be negligible due to our low bubble Reynolds numbers (often below 10, see figure \ref{fig:bubble_cdfs} (right)). Typically, wakes from flow over rigid spheres do not appear until $\Rey_\textrm{b}>20$, while vortex shedding does not occur until about $\Rey_\textrm{b} > 200$ \citep{Clift2005}. More recent studies have also confirmed this fact for flows over spheres in turbulent environments. \citet{Wu1995}, for example, did not observe vortex shedding until $\Rey_\textrm{b}<300$ and note that velocity fluctuations are the same as the surrounding turbulent fluctuations (i.e., without bubbles) for $\Rey_\textrm{b}<300$. This indicates that there should not be severe distortions in the flow field for upstream bubbles, provided their $\Rey_\textrm{b}$ are low. All this allows us to assume that the drag force on our bubbles and the velocity field around them are approximately the same as the case of a single isolated bubble. Substituting these assumptions into equation \ref{x_momentum_balance} yields:
\begin{equation}
  \rho_{\text{v}}V_{\text{b}}\frac{dU_\text{b}}{dt}= (\rho_{\text{l}}-\rho_{\text{v}})gV_{\text{b}} + \frac{1}{2}\rho_{\text{l}}S_{\text{p}}C_{\text{D}}(U_\text{l}-U_\text{b})|U_\text{l}-U_\text{b}| + F_{\text{AM,x}},
  \label{x_momentum_balance_noCL}
\end{equation}
where $S_\text{p}$ is the projected area of the bubble (i.e., $\pi R_{\text{b}}^{2}$), $C_{\text{D}}$ is the drag coefficient and $F_{\text{AM,x}}$ is the added mass force. The drag coefficient can be calculated using the model by \citet{mazzocco_reassessed_2018}, which is the drag coefficient for a solid sphere \citep{zeng_forces_2009} corrected to account for the different boundary conditions between a solid sphere and vapour bubble \citep{legendre_note_1997}:
\begin{equation}
  C_{\text{D}} = 1.13\frac{24}{\Rey_{\text{b}}}\left(1+0.104\Rey_{\text{b}}^{0.753}\right)
  \label{C_D_Mazzocco}
\end{equation}
with the bubble Reynolds number defined as:
\begin{equation}
  \Rey_{\text{b}} =\frac{\rho_{\text{l}}|U_\text{l}-U_\text{b}|D_{\text{b}}}{\mu_{\text{l}}}.
  \label{Rey_b}
\end{equation}
Finally, we note that thermally induced surface tension gradients (i.e., Marangoni forces) are of little concern in our conditions, especially compared to Marangoni forces from surface contaminants \citep{Clift2005}. We note that, if the bubbles were contaminated, then the bubble interface would be a rigid boundary with a no-slip boundary condition, making bubbles behave more like rigid spheres \citep{Harper1967}, in which case the drag coefficient by \citet{zeng_forces_2009} may be used. To elucidate these effects, we compare the \citet{mazzocco_reassessed_2018} and \citet{zeng_forces_2009} correlation in the appendix and observe that the change in bubble sliding predictions are nearly negligible. The liquid velocity at the bubble centroid $U_\text{l}$ is calculated using the \citet{van_driest_turbulent_1956} eddy diffusivity model and McAdams’ friction factor correlation \citep{todreas_nuclear_2011} (more details of the liquid flow analysis are provided in the appendix). The distance from the wall at any given point in time is taken as the bubble radius $R_\text{b}$, because the bubble will remain adhered to the wall during sliding. 
To model the added mass force, we use the approach and results by \citet{favre_updated_2023} and \citet{van_der_geld_dynamics_2009}, which considers the effect of a nearby wall on the flow field surrounding the bubble. The vertical component of the added mass force then becomes:
\begin{equation}
  F_{\text{AM,x}} = C_{\text{AM,x}}\rho_{\text{l}}V_{\text{b}}\left(3\frac{\dot{R}_{\text{b}}}{R_\text{b}}-\frac{dU_\text{b}}{dt}\right)
  \label{F_AM}
\end{equation}
where $C_{\text{AM,x}}=0.636$. The second term in equation \ref{F_AM} represents added inertia due to displacing the liquid as the bubble moves, while the first term actually promotes bubble sliding if the liquid velocity is faster than the vapour bubble, which is expected to be the case. With these closure relations in mind, equation \ref{x_momentum_balance_noCL} can be written as the following nonlinear ordinary differential equation:
\begin{equation}
\begin{gathered}
    (\rho_{\text{v}} + C_{\text{AM,x}}\rho_{\text{l}})V_{\text{b}}\frac{dU_\text{b}}{dt}= \\
    \underbrace{(\rho_{\text{l}}-\rho_{\text{v}})gV_{\text{b}}}_{F_{\text{B}}} + \underbrace{\frac{1}{2}\rho_{\text{l}}S_{\text{p}}C_{\text{D}}(U_\text{l}-U_\text{b})|U_\text{l}-U_\text{b}|}_{F_{\text{D}}} + \underbrace{3C_{\text{AM,x}}\rho_{\text{l}}V_{\text{b}}(U_\text{l}-U_\text{b})\frac{\dot{R}_{\text{b}}}{R_\text{b}}}_{F_{\text{A}}}.
\end{gathered}
  \label{force_balance}
\end{equation}
This equation implies that bubble acceleration will remain positive during the growth phase, and that there are three terms that contribute to bubble sliding. In an effort to simplify this expression, equation \ref{force_balance} can be adjusted to solve for the bubble acceleration:
\begin{equation}
  \begin{gathered}
  \frac{\rho_{\text{v}} + C_{\text{AM,x}}\rho_\text{l}}{\rho_\text{l}}\frac{dU_\text{b}}{dt}=\\
  \left(1-\frac{\rho_\text{v}}{\rho_\text{l}}\right)g + \frac{3}{8}C_\text{D}\frac{|U_\text{l}-U_\text{b}|}{R_\text{b}}(U_\text{l}-U_\text{b})+3C_{\text{AM,x}}(U_\text{l}-U_\text{b})\frac{\dot{R}_{\text{b}}}{R_\text{b}}.
  \end{gathered}
  \label{bubble_eom}
\end{equation}
Next, it will be assumed, and later justified, that the $\Rey_\text{b} < 10$ for most cases, implying: 
\begin{equation}
  C_\text{D} \approx \frac{27.12}{\Rey_\text{b}}
  \label{C_D_simple}
\end{equation}
which yields the following simplification:
\begin{equation}
  \begin{gathered}
  \frac{\rho_{\text{v}} + C_{\text{AM,x}}\rho_\text{l}}{\rho_\text{l}}\frac{dU_\text{b}}{dt}=\\
  \left(1-\frac{\rho_\text{v}}{\rho_\text{l}}\right)g + \frac{3}{16}\frac{27.12\mu_\text{l}}{\rho_{\text{l}}R_\text{b}^{2}}(U_\text{l}-U_\text{b})+3C_{\text{AM,x}}(U_\text{l}-U_\text{b})\frac{\dot{R}_{\text{b}}}{R_\text{b}}.
  \end{gathered}
\end{equation}
Then, by relating the fact that $R_\text{b} = C_{\text{RB}}\sqrt{t}$, we obtain:
\begin{equation}
  \frac{\rho_{\text{v}} + C_{\text{AM,x}}\rho_\text{l}}{\rho_\text{l}}\frac{dU_\text{b}}{dt}=
  \left(1-\frac{\rho_\text{v}}{\rho_\text{l}}\right)g + \left(\frac{81.36\nu_\text{l}}{16C_\text{RB}^{2}}+3C_{\text{AM,x}}\right)\frac{U_\text{l}-U_\text{b}}{t}.
  \label{bubble_eom_simple}
\end{equation}
Finally, the bubble equation of motion is scaled to wall units, where the velocity is scaled by the friction velocity $U_\uptau=\sqrt{\tau_\text{w}/\rho_\text{l}}$, and the length is scaled by $\nu_\text{l}/U_\uptau$. Because the bubble remains adhered to the wall and is relatively small in size during sliding, this scaling procedure is both convenient and physically meaningful. Using $y$ to denote the distance from the wall, where $y=R_\text{b}=C_{\text{RB}}\sqrt{t}$, one obtains:
\begin{equation}
    \frac{dU_\text{b}^{+}}{dy^{+}}=\underbrace{\frac{2\rho_\text{l}}{\rho_{\text{v}} + C_{\text{AM,x}}\rho_\text{l}}\left(\frac{81.36\nu_\text{l}}{16C_\text{RB}^{2}}+3C_{\text{AM,x}}\right)}_{\Pi_{1}}\frac{U_\text{l}^{+}-U_\text{b}^{+}}{y^{+}}+\underbrace{\frac{2g\nu_\text{l}}{U_\uptau^{3}}\frac{1-\rho_\text{v}/\rho_\text{l}}{C_{\text{AM,x}}+\rho_\text{v}/\rho_\text{l}} \sqrt{\frac{\nu_\text{l}}{C_\text{RB}^{2}}}}_{\Pi_{2}}.
    \label{bubble_eom_wall_units}
\end{equation}
Because the liquid velocity profile is strictly a function of $y^+$, the differential equation can be solved analytically with an integrating factor, yielding the exact solution:
\begin{equation}
    U_\text{b}^{+}=\frac{\int_{0}^{y^{+}}\Pi_{1}y^{+^{\Pi_{1}-1}}U_\text{l}^{+}+\Pi_{2}y^{+^{\Pi_1-1}} dy^{+}}{y^{+^{\Pi_{1}}}}
    \label{ub_int_pi2}
\end{equation}
which can be readily solved using a conventional eddy-diffusivity model. It is noted that the gravitational component in this equation is very small for the operating conditions here, i.e., $\Pi_2 \ll1$, and so it is neglected for the sake of simplicity: 
\begin{equation}
    U_\text{b}^{+}=\frac{\int_{0}^{y^{+}}\Pi_{1}y^{+^{\Pi_{1}-1}}U_\text{l}^{+}dy^{+}}{y^{+^{\Pi_{1}}}}
    \label{ub_int}
\end{equation}
which results in a function that is dependent on just one non-dimensional number, $\Pi_1$. A phenomenological interpretation of $\Pi_1$ can be understood by considering how drag, added mass and inertial forces change as bubbles get smaller. Precisely, smaller bubbles will have less inertia but a larger area-to-volume ratio and drag coefficient, as shown in equation \ref{bubble_eom}. This results in larger bubble acceleration, which makes them more rapidly approach the velocity of the surrounding liquid. Therefore, $\Pi_1$ as defined in equation \ref{bubble_eom_wall_units} can be viewed as a ratio of drag (first term in parentheses) and added mass forces (second term in parentheses) to bubble and liquid inertial forces (first and second terms in the denominator outside the parentheses), where bubbles with larger values of $\Pi_1$ will more closely match the liquid velocity, all else being equal. The simplified analytical model for bubble velocity (i.e., equation \ref{ub_int}) is compared against all measured bubble velocity data in figure \ref{fig:law_of_wall} (black solid line), where the experimentally measured bubble velocities and radii (represented as distance from the wall) for all operating conditions are plotted in wall units. In plotting equation \ref{ub_int}, the quantity $\Pi_1$ is held constant at 5.8 as it falls within the typical spread of measured $\Pi_1$ values (3.3 - 8.3). We also plot the theoretical profile for the liquid velocity (black dashed line) according to \citet{van_driest_turbulent_1956}, which in the limiting case of short distance from the wall is linear, whereas at far distances from the wall is logarithmic. Notably, equation \ref{ub_int} can be further simplified in a few special cases. When the bubble always remains in the viscous sublayer while sliding, the solution becomes (shown as a red dotted line in figure \ref{fig:law_of_wall}):
\begin{figure}
  \centerline{\includegraphics[width=\textwidth]{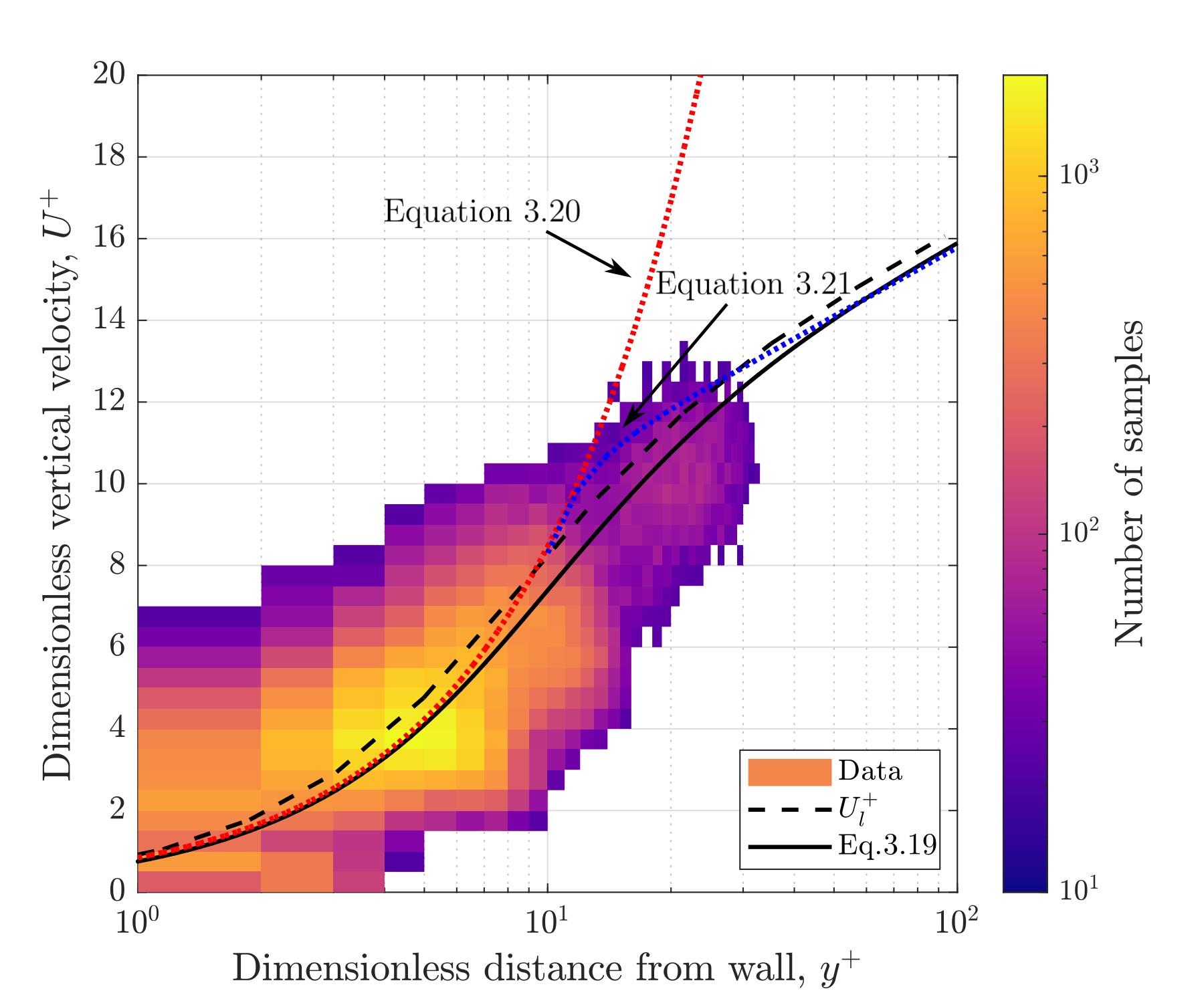}}
  \caption{Experimentally measured bubble velocities and distances from wall across all operating conditions, with theoretical predictions overlayed.}
\label{fig:law_of_wall}
\end{figure}
\begin{equation}
    U_\text{b}^{+}=\frac{\Pi_1}{\Pi_1+1}y^+=\frac{\Pi_1}{\Pi_1+1}U_\text{l}^+
    \label{ub_lin}
\end{equation}
It can be seen from the data that the measured bubble velocity follows this linear trend (i.e., equation 20) up until values of $y^+ \sim 10$ are reached. 
Once the bubble leaves the viscous sublayer and gets into the turbulent core, where $U_\text{l}^{+}=\frac{1}{\kappa}\ln y^+ + C^+$, then the bubble velocity may be expressed as (shown as a blue dotted line in figure \ref{fig:law_of_wall}):
\begin{equation}
   U_\text{b}^{+}=U_\text{l}^{+} + \frac{\Pi_1}{\Pi_1+1}\frac{y_{\text{n}}^{+^{\Pi_1+1}}}{y^{+^{\Pi_1}}}-C^{+}\frac{y_{\text{n}}^{+^{\Pi_1}}}{y^{+^{\Pi_1}}}-\frac{1}{\kappa}\left(\frac{y_{\text{n}}^{+^{\Pi_1}}\ln y_{\text{n}}^+}{y^{+^{\Pi_1}}}+\frac{1}{\Pi_1} - \frac{y_{\text{n}}^{+^{\Pi_1}}}{\Pi_{1}y^{+^{\Pi_1}}} \right)
    \label{ub_log}
\end{equation}
where $\kappa=0.41$, $C^+=5.0$ and $y_\text{n}^+$ is the location at which the log-law and linear velocity profile intersect ($\sim 10.80$). The slip ratio (i.e., ratio of bubble velocity to liquid velocity) is a constant value less than unity when the bubble is in the sublayer (see equation \ref{ub_lin}). Far from the wall, as $y^+$ approaches infinity, equation \ref{ub_log} becomes:
\begin{equation}
     U_\text{b}^{+}= U_\text{l}^{+}-\frac{1}{\kappa\Pi_1}
\end{equation}
and thus, the slip ratio tends to unity. Overall, the measured bubble velocities can be explained by the analytical solutions proposed here, further suggesting that high-pressure bubble dynamics can be properly accounted for without considering other forces related to bubble asymmetry, which are difficult to measure in practice and add significant complexity to bubble departure modeling.  We acknowledge that this analysis assumes that the velocity profile is unaffected by the presence of bubbles, as mentioned before, and re-emphasize that this concern is alleviated in the limiting case of small bubbles and $U_{\text{b}}\sim U_{\text{l}}$ (i.e., low $\Rey_\textrm{b}$ and negligible slip), which is the case in high-pressure flow boiling conditions.
The analytic bubble velocity expressions can also be used to calculate the vertical displacement of the bubble. For example, if the bubble remains in the viscous sublayer (i.e., $U_\text{l}^+=y^+$), then from equations \ref{R_CRB} and \ref{ub_lin}, we get:
\begin{equation}
    x_\text{b}(t)=\frac{2}{3}\frac{\Pi_1}{\Pi_1+1}\frac{C_\text{RB}^{2}U_\uptau^2}{\nu_\text{l}}t^{3/2}
    \label{xb_lin}
\end{equation}
Otherwise, the bubble displacement can be computed using a specific eddy-diffusivity model and relatively simple numerical integration:
\begin{equation}
    x_\text{b}(t)=\int_{0}^{t}U_\text{b}^{+}U_\uptau dt
    \label{xb_int}
\end{equation}
The combination of equations \ref{ub_lin}, \ref{ub_log} and \ref{xb_int} (taking $y_\text{n}^+=10.8$) results in a completely analytical expression for the bubble trajectory during sliding. Note that, in both equations \ref{xb_lin} and \ref{xb_int}, we assumed that a bubble begins to slide as soon as it nucleates, an idea supported by the optical measurements in figure \ref{fig:PD_highp}. To validate this approach, results from the numerical integration of the exact differential equation (equation \ref{bubble_eom}), analytical solution (equations \ref{ub_lin}, \ref{ub_log} and \ref{xb_int}) and experimentally measured bubble trajectories are shown in figures \ref{fig:sliding_lowG}-\ref{fig:sliding_highG}. Like the bubble growth predictions, variations in the bubble growth fitting parameter $C_\text{RB}$ can also explain most of the statistical spread in the bubble trajectory data. Unlike the bubble growth results, there is a clear dependence on the mass flux due to the derivations and analysis described above. Increasing pressure tends to slow down the bubble because of the decrease in $C_\text{RB}$ (e.g., see equation \ref{xb_lin}). Finally, one can observe the excellent agreement between the differential equation solution and the simplified analytical solution, even though simplified drag coefficients were used, and buoyancy was neglected. A sensitivity study on the impact of drag and buoyancy was performed, and it was found, and shown in the appendix, that the simplification of the drag coefficient had a larger, yet insignificant impact on the model predictions.
\begin{figure}
  \centerline{\includegraphics[width=\textwidth]{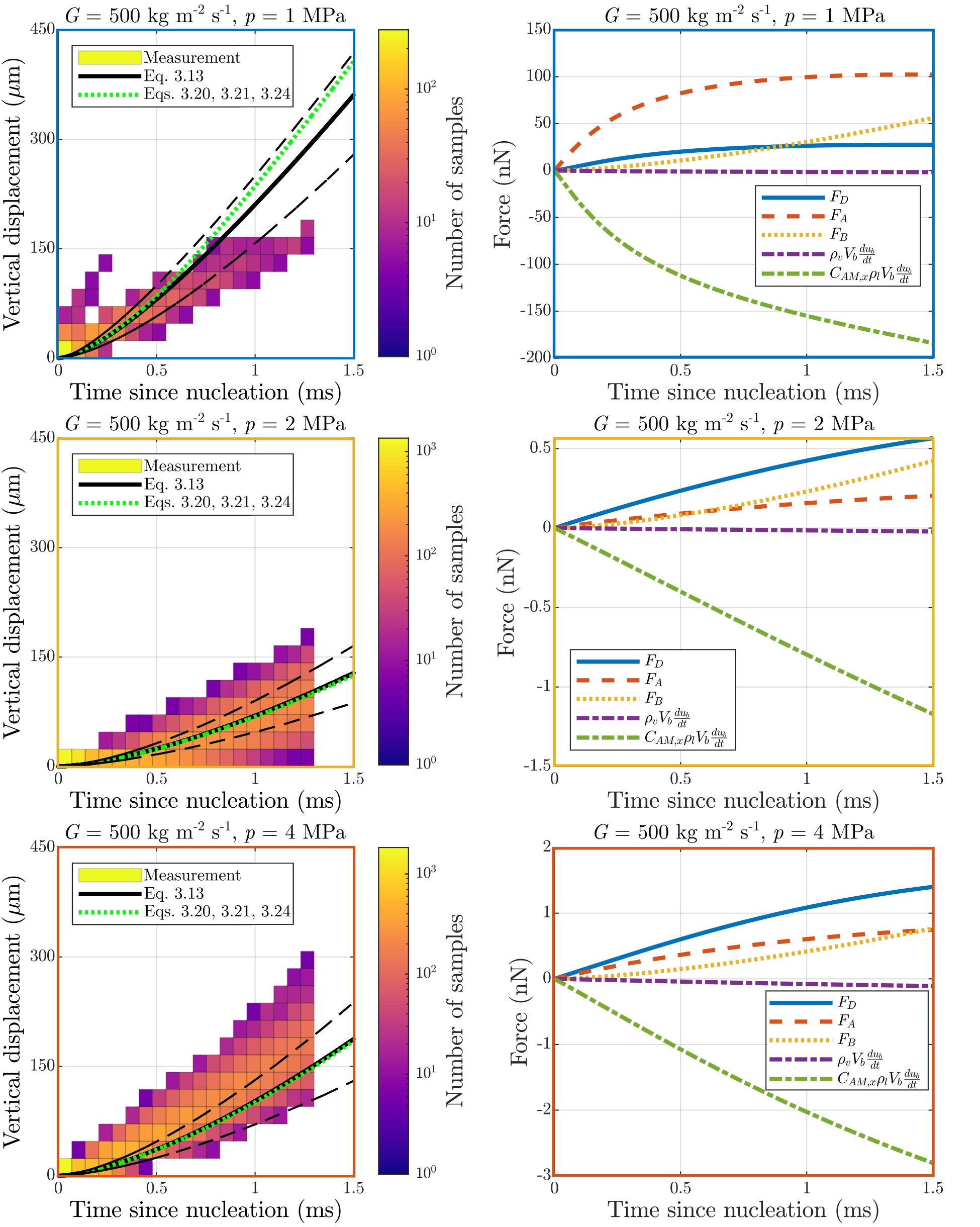}}
  \caption{Distance bubble travels (left) and forces acting on bubble (right) while sliding at a mass flux of 500 kg m$^{-2}$ s$^{-1}$.}
\label{fig:sliding_lowG}
\end{figure}
\begin{figure}
  \centerline{\includegraphics[width=\textwidth]{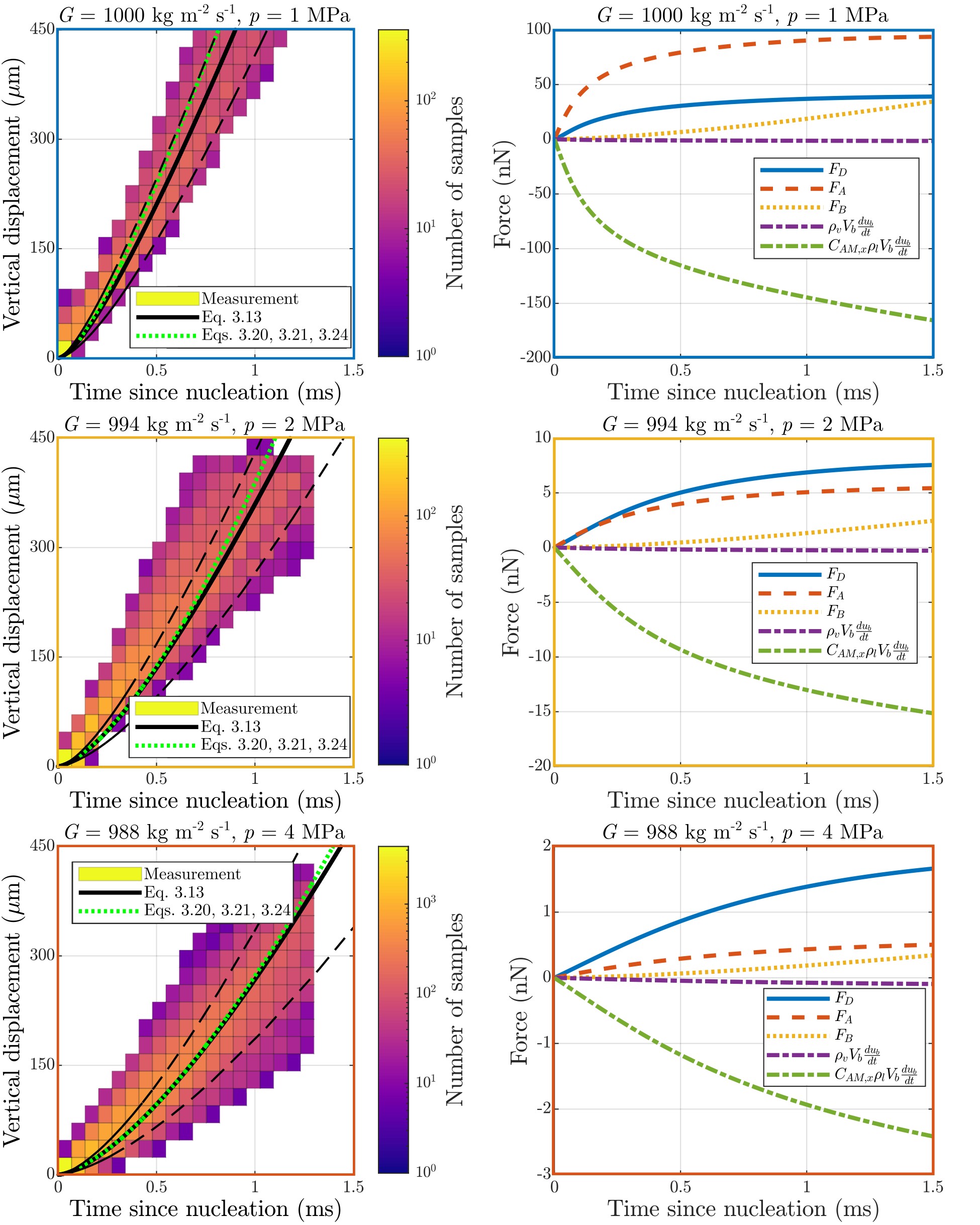}}
  \caption{Distance bubble travels (left) and forces acting on bubble (right) while sliding at a mass flux of 1000 kg m$^{-2}$ s$^{-1}$.}
\label{fig:sliding_medG}
\end{figure}
\begin{figure}
  \centerline{\includegraphics[width=\textwidth]{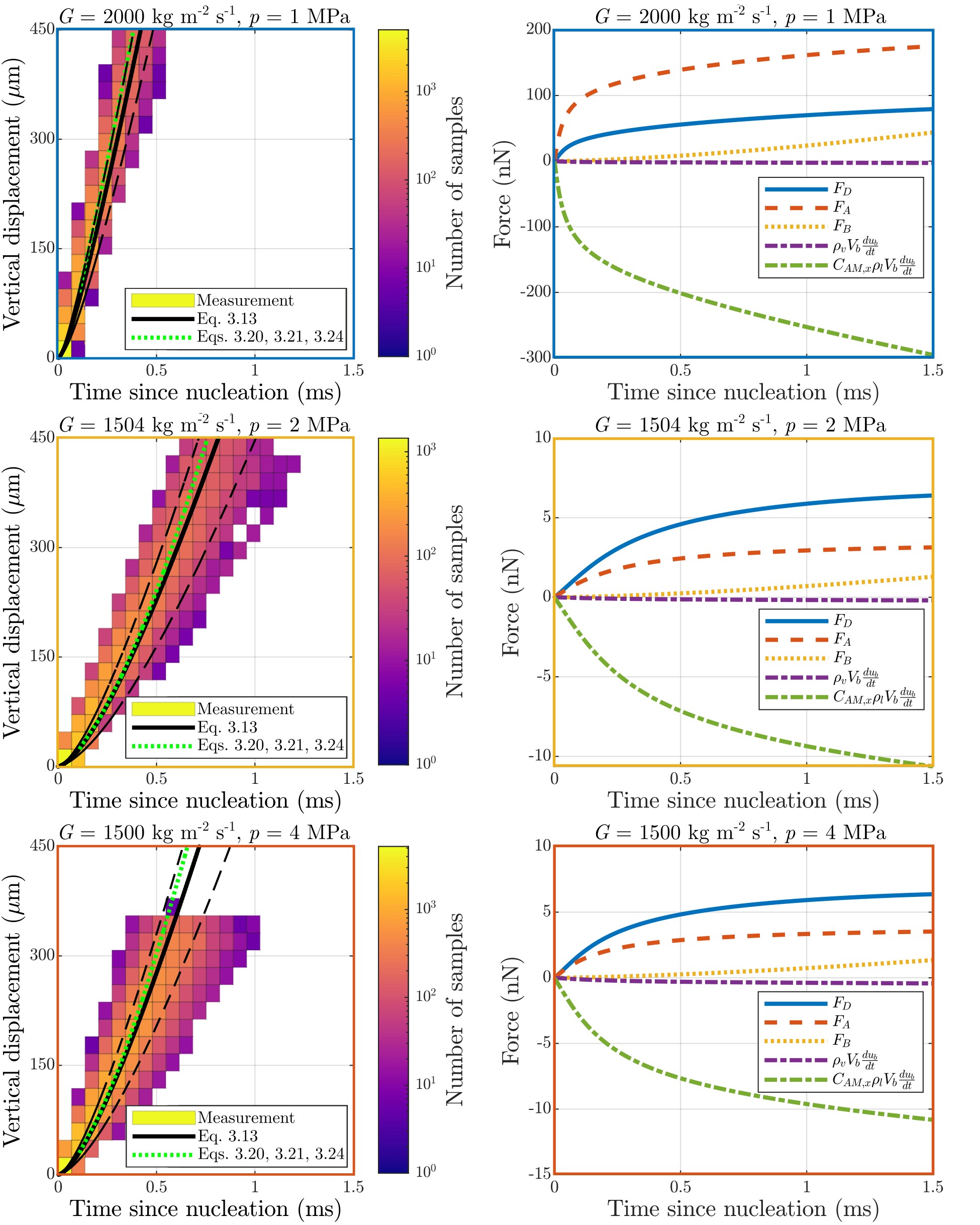}}
  \caption{Distance bubble travels (left) and forces acting on bubble (right) while sliding at the highest mass fluxes tested (2000 kg m$^{-2}$ s$^{-1}$ at 1 MPa and 1500 kg m$^{-2}$ s$^{-1}$ for 2-4 MPa).}
\label{fig:sliding_highG}
\end{figure}
Figures \ref{fig:sliding_lowG}-\ref{fig:sliding_highG} also show the magnitude of the forces acting on the bubble during sliding. At a fixed pressure, when $C_\text{RB}$ is effectively constant, increasing mass flux increases the magnitude of the drag force. This is of course expected, as viscous drag forces are approximately proportional to the velocity of the flow (noting that the slip ratio is approximately uniform), which will obviously increase with mass flux, both because of the larger mean fluid velocity and smaller viscous sublayer thickness. At a fixed mass flux, increasing pressure, i.e., smaller $C_\text{RB}$, tends to increase the role of drag forces, as suggested by equation \ref{bubble_eom_simple}.

In passing, we note that the bubble sliding phenomenon can be conceptualized and modeled as a small particle moving in a liquid medium, leading to simple and convenient equations of motion that are strongly supported by experimental measurements. The major conceptual difference between the two phenomena is the fact that the vapour bubble is growing with time. If the bubble remains at a fixed size, then the equation of motion is simply:
\begin{equation}
    \frac{dU_\text{b}}{dt}+\frac{9\nu_\text{l}}{2R_\text{b}^{2}}U_\text{b}=\frac{9\nu_\text{l}}{2R_\text{b}^{2}}U_\text{l}
    \label{particle_eom}
\end{equation}
This first-order differential equation predicts that the particle velocity reaches the liquid velocity after $\sim 1.1R_\text{b}^2/\nu_\text{l}$. In the case of the growing vapour bubble, there is always some nonzero slip that causes the bubble to never exactly reach the liquid velocity, unless the quantity $\Pi_1$ is very large, in which case the bubble velocity is essentially the liquid velocity for all practical purposes.
\subsection{Bubble departure}
\begin{figure}
  \centerline{\includegraphics[width=\textwidth]{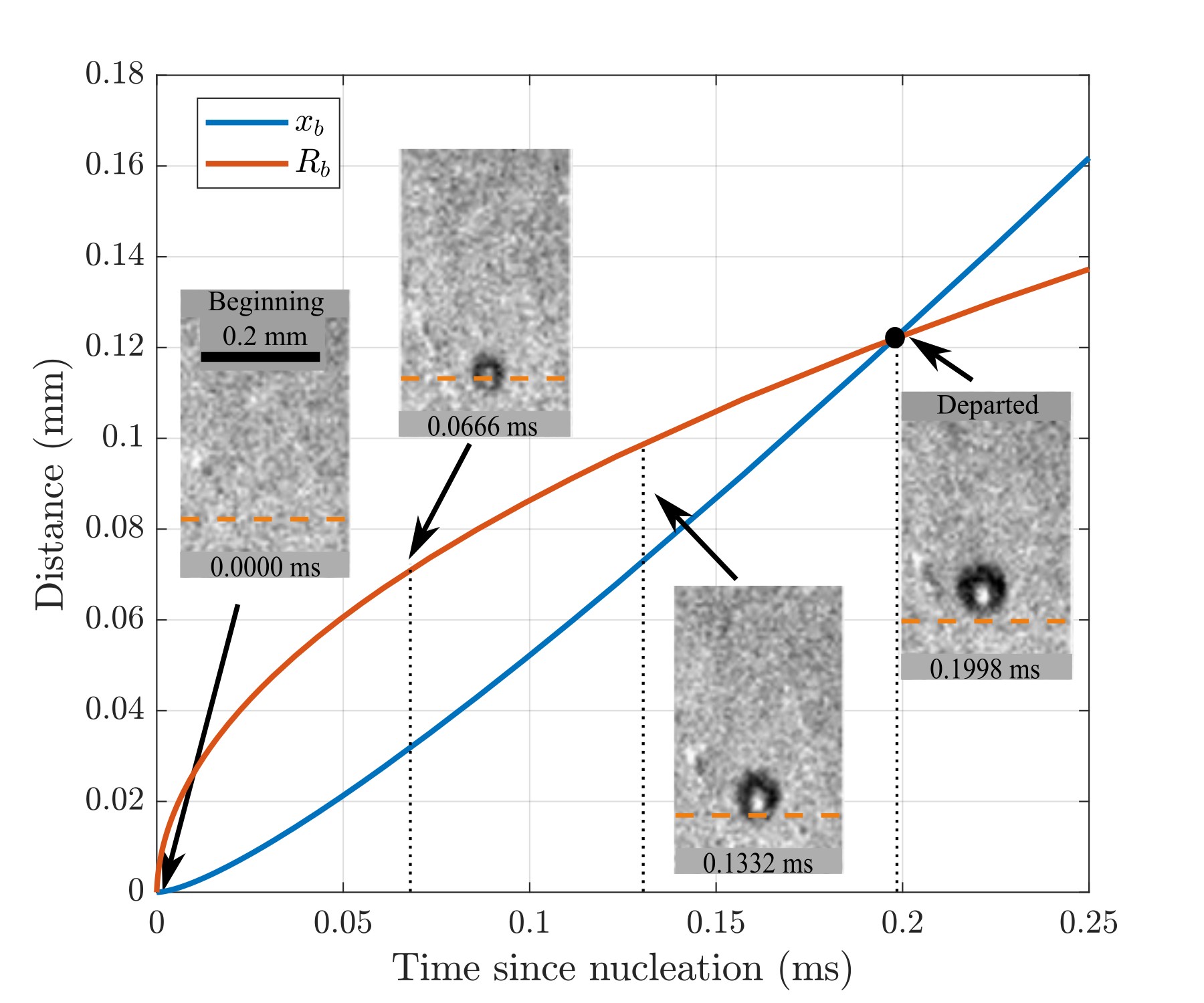}}
  \caption{Proposed departure criterion illustrated with departure data at 1 MPa, with a mass flux and heat flux of 1000 kg m$^{-2}$ s$^{-1}$ and 500 kW m$^{-2}$, respectively. Note that the scales for the graph and images are different.}
\label{fig:Dd_criterion}
\end{figure}

\begin{figure}
  \centerline{\includegraphics[width=\textwidth]{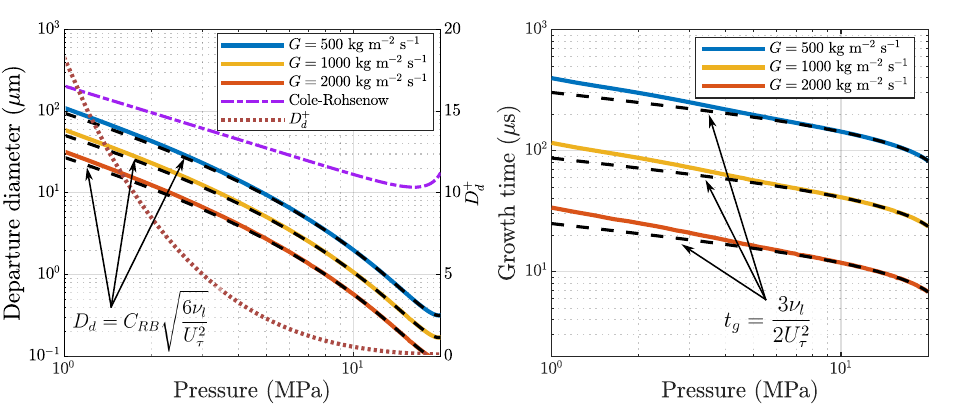}}
  \caption{Departure diameter (left) and growth time (right) predictions as a function of pressure for various mass fluxes. Solid lines show predictions from equation \ref{bubble_eom}, while the dashed lines show predictions from equations \ref{tg_limit} and \ref{Dd_limit} For the sake of illustration, $C_\text{RB}=B/2$ and $q''=500$ kW m$^{-2}$.}
\label{fig:departure_trends}
\end{figure}

The observations of the high-speed video images and the analysis in the previous section support the hypothesis that bubble sliding begins as soon as the bubble nucleates. This interesting finding leaves us with an open question: how to define bubble departure diameter and growth time? We may define the bubble departure diameter as the equivalent diameter that the bubble has when its optical footprint stops covering its original nucleation site, allowing for a new bubble to nucleate. This is defined to be consistent with the departure diameter and growth time definitions adopted in heat flux partitioning models. This criterion is illustrated in figure \ref{fig:Dd_criterion}, where the bubble is first covering the nucleation site and later moves past it after $\sim$0.2 ms with a diameter of 0.1 mm, which define grow time and departure diameter, respectively. For a spherical bubble, the departure criterion can be represented mathematically as the time at which:
\begin{equation}
    x_\text{b}(t)=\int_{0}^{t}U_\text{b}(t)dt=R_\text{b}(t)
    \label{Dd_criterion}
\end{equation}
Equation \ref{Dd_criterion} can be readily solved analytically. In the case where the bubble departs from the nucleation site while inside the viscous sublayer (which will be shown to be the case the vast majority of the time), the growth time and departure diameter can be evaluated as
\begin{equation}
    t_\text{g}=\frac{3}{2}\frac{\Pi_1+1}{\Pi_1}\frac{\nu_\text{l}}{U_\uptau^2}
    \label{tg_lin}
\end{equation}
and
\begin{equation}
    D_\text{d}=C_\text{RB}\sqrt{\frac{6(\Pi_1+1)}{\Pi_1}\frac{\nu_\text{l}}{U_\uptau^2}},
    \label{Dd_lin}
\end{equation}
respectively. In the limit where $\Pi_1\gg 1$, $U_\text{b}=U_\text{l}$,
\begin{equation}
    t_\text{g}=\frac{3}{2}\frac{\nu_\text{l}}{U_\uptau^2}
    \label{tg_limit}
\end{equation}
\begin{equation}
    D_\text{d}=C_\text{RB}\sqrt{\frac{6\nu_\text{l}}{U_\uptau^2}},
    \label{Dd_limit}
\end{equation}
which provide extremely simple expressions for the departure diameter and growth time. To demonstrate that these simple formulae can be sufficiently accurate, the departure diameter and growth time determined numerically using equation \ref{Dd_criterion} and the complete force balance model (i.e., equation \ref{bubble_eom}) are compared against equations \ref{tg_limit} and \ref{Dd_limit}. Results are shown in figure \ref{fig:departure_trends}, where it is immediately clear that, for most pressures above 1 MPa, the simple analytical expressions for departure diameter and growth time compare extremely well with the more complex force balance model. Note that equations \ref{tg_lin} and \ref{Dd_lin} would lie in between the predictions of the force balance and equations \ref{tg_limit} and \ref{Dd_limit} and are not shown in figure \ref{fig:departure_trends} for the sake of clarity.

It can also be seen that equations \ref{tg_limit} and \ref{Dd_limit} consistently underpredict the results at lower pressures. Assuming the bubble remains in the viscous sublayer, this can be understood by considering the effect pressure has on the quantity $\Pi_1$ and consequently, equations \ref{tg_lin} and \ref{Dd_lin}. Because decreasing pressure increases $C_\text{RB}$, $\Pi_1$ also decreases (see equation \ref{bubble_eom_wall_units}). Since $\Pi_1$ is directly related to the vapour-liquid slip velocity ratio (equation \ref{ub_lin}), decreasing $\Pi_1$ increases the relative slip between the two phases, making the assumption inherent to equations \ref{tg_limit} and \ref{Dd_limit} (i.e., $U_\text{b}=U_\text{l}$) less accurate.  It must also be confirmed that the bubble remains in the sublayer during sliding. Figure \ref{fig:departure_trends} shows that the bubble (see curve labelled  $D_\text{d}^+$) remains in the viscous sublayer beyond a pressure of about 2 MPa, and is never far inside the turbulent core. Therefore, both assumptions (i.e., that the bubble remains in the viscous sublayer and that $U_\text{b}=U_\text{l}$) are more accurate at elevated pressures.

Based on equations \ref{tg_limit} and \ref{Dd_limit}, one can easily interpret how different operating conditions affect bubble departure. For example, it can be readily deduced that the departure diameter is approximately inversely proportional to the mass flux ($\propto G^{0.9}$) and directly proportional to $C_\text{RB}$, while the growth time is inversely proportional to $\propto G^{1.8}$ and does not at all depend on how fast the bubble grows. This observation can be deduced from the fact that the speed of the bubble is directly related to its size while in the viscous sublayer, which makes the time at which the bubble departs independent of its growth rate.

Interestingly, it can also be observed that all departure diameters shown in figure \ref{fig:departure_trends} lie below the Cole-Rohsenow predictions for pool boiling conditions (purple dash-dotted line figure \ref{fig:departure_trends}). This is expected due to the presence of drag and added mass forces promoting bubble sliding. However, it must be noted that this finding should not be extrapolated to pressures lower than what was investigated in this study, as bubble deformation effects, due to their size, will create non-negligible surface tension forces \citep{favre_updated_2023} that make our force balance model and analytical solution invalid. 

\begin{figure}
  \centerline{\includegraphics[width=\textwidth]{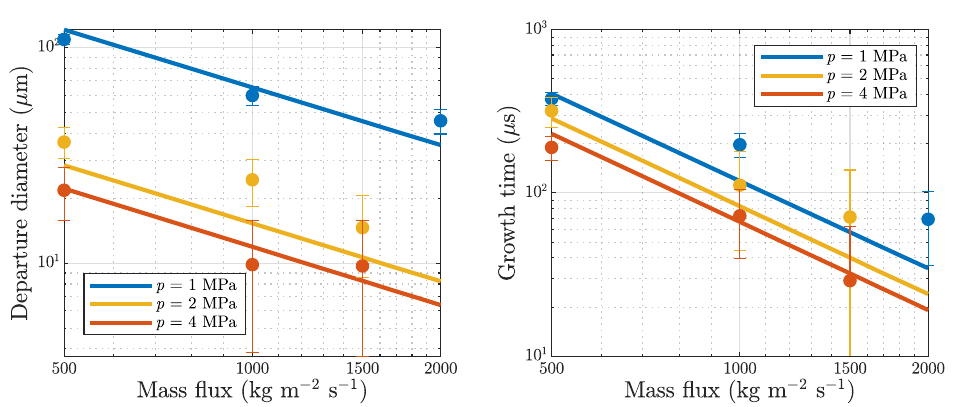}}
  \caption{Departure diameter (left) and growth time (right) predictions and measurements as a function of mass flux for various pressures.}
\label{fig:departure_error}
\end{figure}

Equation \ref{Dd_criterion} can also be used to measure the departure diameter and growth time experimentally. This is achieved by our bubble tracking algorithm, where the intersection of the two curves in figure \ref{fig:Dd_criterion} is determined using linear interpolation. A comparison between average experimental measurements and theoretical predictions is shown in figure \ref{fig:departure_error}.
The error bars in the departure diameter and growth time measurements come mostly from the spatial ($\sim$5 – 6 $\mu$m) and temporal (33 – 67 $\mu$s) resolution of the optical technique. In general, departure diameter predictions improve as pressure increases, as expected based on the model assumptions. It can also be seen that the growth time predictions are also reasonably accurate at higher pressures, but are not quite as precise as the departure diameter predictions. This happens because bubbles may not be precisely tracked the instant they are nucleated. Therefore, the experimentally measured growth time becomes less accurate the faster the bubbles depart. This is not the case for departure diameter, because the measurement is only taken when the detected bubble leaves the nucleation site area. In general, it is concluded that the model is in agreement with the optical measurements and their associated uncertainty. It is also clear that the measured departure diameters are indeed far smaller than what has been reported in past literature studies (see figure \ref{fig:Dd_literature}), as was expected. Even though there is no universal definition of departure diameter, our solution offers a physically sound view of the bubble departure process in high-pressure flow boiling compared to approaches previously used in literature. Additionally, our analytical solution may offer much more tractable and computationally efficient ways to evaluate bubble departure and growth time compared to numerically solving complex equations of motion. In passing, we note that the assumption of departure diameter approaching zero as pressure increases has also been made in recent CFD flow boiling models \citep{kommajosyula_development_2020}, which is consistent with our mathematical models and experimental measurements.

\begin{figure}
  \centerline{\includegraphics[width=\textwidth]{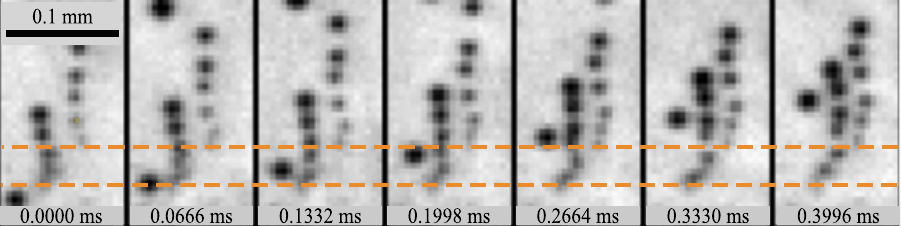}}
  \caption{Bubble columns appearing from two nucleation sites in 10 ºC subcooled flow boiling at 7.6 MPa with a mass flux and heat flux of 1000 kg m$^{-2}$ s$^{-1}$ and 220 kW m$^{-2}$, respectively.}
\label{fig:76bar}
\end{figure}

Moreover, equations \ref{tg_limit} and \ref{Dd_limit} can provide a simple and meaningful to predict the effect of pressure and flow rate on the bubble departure process. As an example, we can use them to evaluate what would be the spatial and temporal resolution needed to measure these quantities at PWR conditions ($\sim$15 MPa and $>$ 2000 kg m$^{-2}$ s$^{-1}$). The departure diameter would be in the order of 100 nm and the growth time less than 10 $\mu$s. Such a small diameter cannot be resolved with the optical technique used in this study, which is limited to about 6 $\mu$m. It may be possible to achieve sub-micron spatial resolution with optical microscopes, but such devices will have to be positioned a few millimeters away from the boiling surface, which makes the fabrication of a seal capable of withstanding $\sim$15 MPa increasingly difficult. In fact, flow boiling at a pressure of 7.6 MPa was attempted in the present study (shown in figure \ref{fig:76bar}), but the uncertainty in the optical measurements was quite large. At a pressure of 7.6 MPa and mass flux of 1000 kg m$^{-2}$ s$^{-1}$, departing bubbles appear to be a few pixels or even less than one pixel. Based on the images shown in figure \ref{fig:76bar}, the departure diameter appears to be approximately two to three microns. If we refer to figure \ref{fig:departure_trends}, it can be seen that the model predicts a departure diameter of about 2 microns at a pressure of 7.6 MPa and 1000 kg m$^{-2}$ s$^{-1}$ mass flux, which indicates that the present model and measurements are approximately in agreement, but will need to be validated with future measurements. Based on these findings and limitations, future studies that focus on bubble departure in high pressure flow boiling applications will either have to use a more advanced optical setup with higher spatial resolution or make use of light with shorter wavelengths (e.g., ultraviolet or x-rays, provided that these techniques have sufficient temporal resolution).

\section{Conclusions}\label{sec:conclusions}
We conducted an experimental investigation on the growth, departure and sliding of vapour bubble in high-pressure flow boiling using high-resolution optical techniques. With these techniques we could track bubble shape, size and position as bubbles nucleate and slide on top of the heated surface. In our experimental conditions, the radii of bubbles approximately grow with the square-root of time, which hints towards a heat-diffusion controlled process. We also observed that the bubble shape is essentially spherical throughout the growth and sliding process, as their small size makes surface tension forces dominate over inertial, viscous and gravitational forces.

These observations allowed us to simplify the bubble equation of motion and constitutive laws, e.g., for the drag coefficient, to obtain non-dimensional expressions for the bubble trajectory in the direction of the flow. Specifically, we identified a non-dimensional number, $\Pi_1$, that relates the bubble velocity to the liquid velocity and depends only on bubble growth rate and physical properties. In the viscous sublayer, the slip ratio is fixed, while in the turbulent flow region, the relative velocity between the bubble and the liquid is fixed and the slip ratio tends to unity. Notably, in high-pressure flow boiling, as the bubble growth rate gets smaller, $\Pi_1$ tends to infinity, making the bubble velocity essentially equal to the liquid velocity. This observation yields analytic expressions for bubble departure diameter and growth time that only depend on bubble growth rate, liquid kinematic viscosity and shear velocity.  This offers a much more tractable and computationally effective way to evaluate these quantities compared to numerically solving complex equations of motion. For this reason, these expressions can readily and efficiently be used in two-phase CFD modeling frameworks.

Unfortunately, due to the of lack wall and local fluid temperature measurements, we could not formulate a comprehensive mechanistic model of the bubble growth. In future works, we will focus on implementing these measurements and study the effect of subcooling and nucleation temperature on bubble growth in high-pressure flow boiling. Future work may also focus on surface effects and bubble coalesce during sliding, as the formation of large vapour bubbles that lift off the surface becomes increasingly important as heat fluxes approach the CHF limit.


\backsection[Acknowledgements]{We thank J. Buongiorno, E. Baglietto and A. Roccon for many inspiring discussions and comments on our work.}

\backsection[Funding]{This work was performed with the funding support of the US Department of Energy 2021 Distinguished Early Career Award (award number DE-NE0009321) and support of the Consortium for Advanced Simulations of Light Water Reactors under the US Department of Energy Contract No. DEAC05-00OR22725.}

\backsection[Declaration of interests]{The authors report no conflict of interest.}

\backsection[Data availability statement]{The data that support the findings of this study are available from the corresponding author upon reasonable request.}

\appendix

\section{}\label{appA}
\subsection{High-pressure flow boiling test facility}
\begin{figure}
  \centerline{\includegraphics[width=\textwidth]{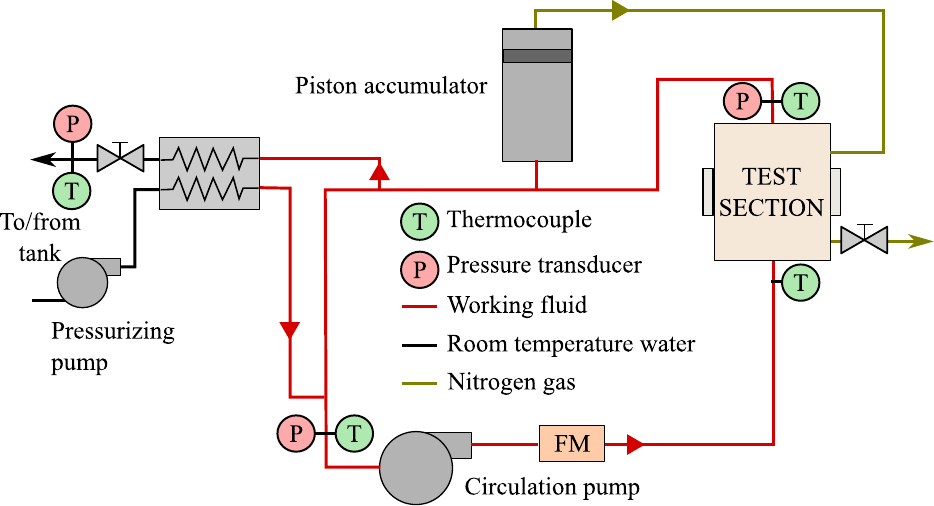}}
  \caption{Schematic of the flow boiling test facility.}
\label{fig:test_facility}
\end{figure}

A schematic of the flow boiling test facility, including the required instrumentation, is shown in figure \ref{fig:test_facility}. The facility is designed to operate at 15.5 MPa, and it is possible to adjust flow rate and water temperature to duplicate pressurized water reactors (PWR) flow conditions. Essentially, a circulation pump is used in the working fluid (high pressure, high temperature) loop to establish the required mass flux through the test section, which is measured by a venturi-type flow meter (FM). The temperature of the flow loop upstream of the test section is raised and controlled by a series of tape heaters wrapped around the working fluid tubing. To control the system pressure, a pressurizing pump is used to take room temperature water from a tank and injects it into the system. Simultaneously, a controlled discharge of water is established with a backpressure regulator. Both these mechanisms create a “feed-and-bleed” dynamic. Both streams are also sent through a heat exchanger, which cools the bleed stream and preheats the feed stream. A piston accumulator is installed near the test section outlet to minimize the pressure difference between the sapphire windows that separate the working fluid from the test section interior. Deionized water was used as the working fluid for all experiments. Before each test, argon gas was sparged through the water in the tank to remove oxygen. The average experimental uncertainties of the test facility measurements are shown in table \ref{tab:experiment_uncertainties}. 

\begin{table}
  \begin{center}
\def~{\hphantom{0}}
  \begin{tabular}{lc}
      \textit{Parameter}         & \textit{Uncertainty}\\ [6pt]
      System pressure   & $\pm 0.02$ MPa\\ [2pt]
      Subcooling        & $\pm 1.1$ K\\[2pt]
      Mass flux         & $\pm 20$ kg m$^{-2}$ s$^{-1}$\\[2pt]
      Heat flux         & $\pm 100$ kW m$^{-2}$
  \end{tabular}
  \caption{Uncertainty of measurements from high-pressure flow boiling test facility.}
  \label{tab:experiment_uncertainties}
  \end{center}
\end{table}

\subsection{Bubble tracking algorithm}
\begin{figure}
  \centerline{\includegraphics[width=0.75\textwidth]{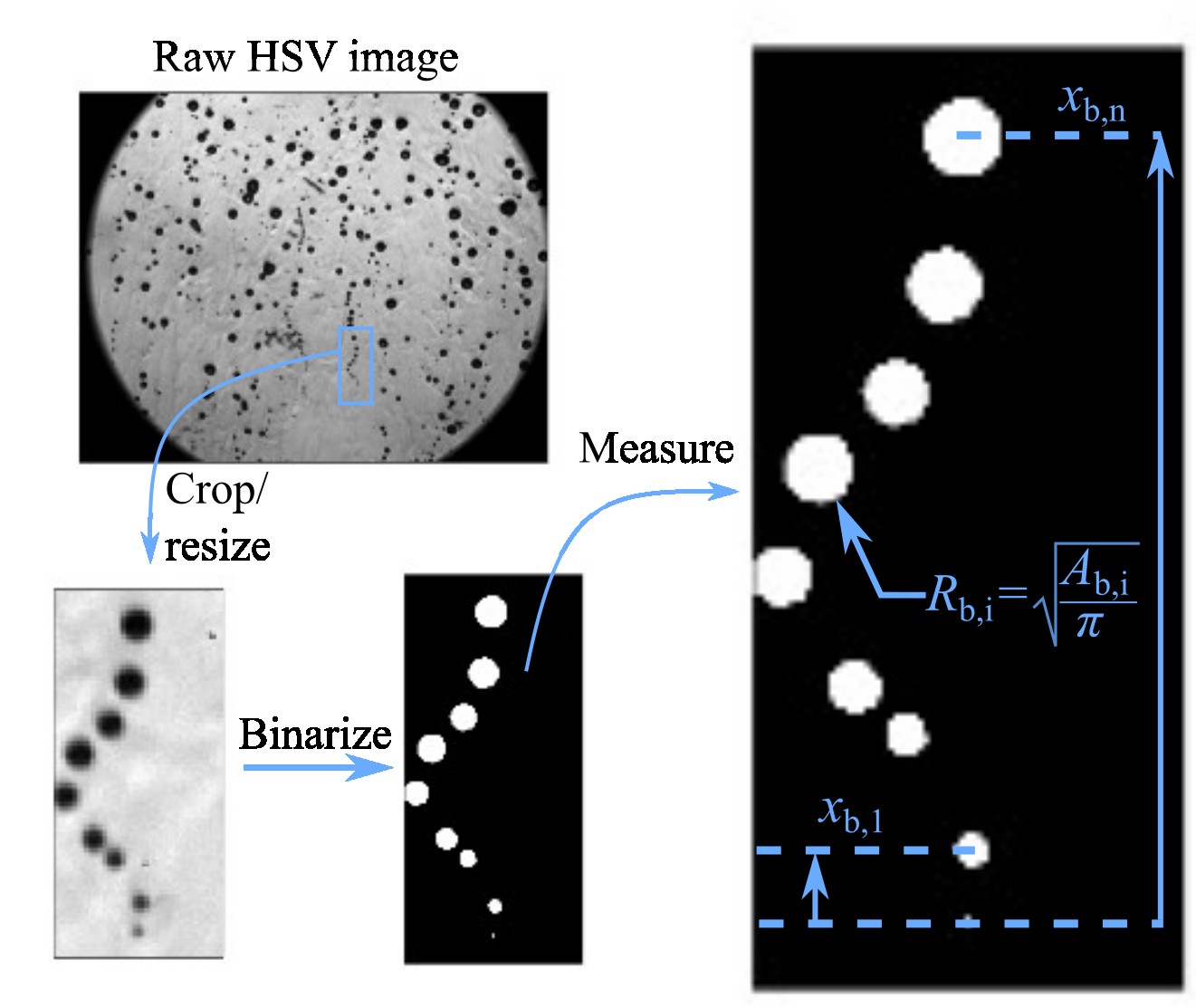}}
  \caption{ Flowchart for bubble tracking algorithm.}
\label{fig:flowchart}
\end{figure}

\begin{figure}
  \centerline{\includegraphics[width=0.8\textwidth]{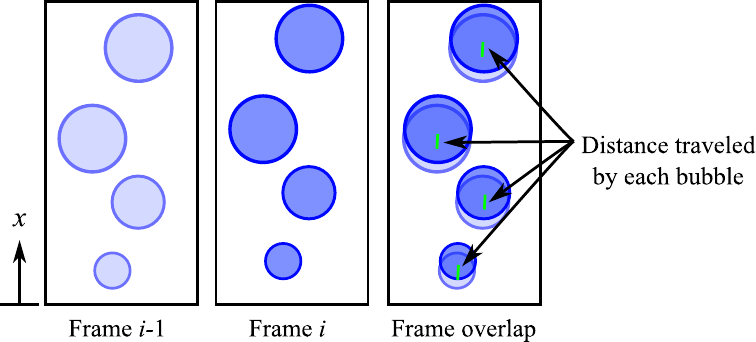}}
  \caption{Illustration of the procedure that used to correlate bubble positions between two consecutive frames.}
\label{fig:tracking}
\end{figure}

The algorithm to track the size and trajectory of bubbles was developed in MATLAB®. Figure \ref{fig:flowchart} shows a flowchart depicting the key steps within the algorithm. Before the algorithm begins, nucleation sites that consistently produce bubbles are manually identified. Once nucleation site regions of interest are identified, the algorithm begins by cropping the raw image to only show the region surrounding a nucleation site. Then, each cropped image, or frame, $i$ of a video is resized by a factor of five to improve the detection of the bubble perimeter. Once the region of interest has been enlarged, each image is binarized and a series of built-in MATLAB® functions are used to calculate the number of white regions (i.e., detected bubbles) and their respective centroid and area.
To track the bubbles over the course of a video, bubbles detected in frame $i$ of the video are compared with the bubbles detected in the preceding frame $i-1$. This comparison is illustrated in figure \ref{fig:tracking} and briefly described here. Consider a bubble in frame $i-1$ and compare its centroid with the centroid of all bubbles in frame $i$. The pair of centroids that yields the smallest Euclidean distance is considered to be the same bubble, provided the trajectory of the bubble is positive in the vertical-direction; if the trajectory is negative in the vertical direction, then this pair is discarded. This procedure continues for all bubbles in frame $i-1$. Once all these bubbles are paired, then any remaining unpaired bubbles in frame $i$ are either ignored or recorded as newly nucleated bubbles, depending on their Euclidean distance from the nucleation site. If the distance is close enough, then the bubble is considered to be newly nucleated; otherwise, the bubble is discarded.
It should be noted that the algorithm requires videos recorded at a high frame rate relative to the bubble motion so that bubble tracking can be achieved. Additionally, there must be little interaction between bubbles since effects such as bubble coalescence are not considered. For this reason, only low heat flux data are reported in the present study, as flow-boiling at higher heat fluxes will produce too many bubbles that will undoubtedly interact with each other.

\subsection{Image processing uncertainty analysis}
\begin{figure}
  \centerline{\includegraphics[width=\textwidth]{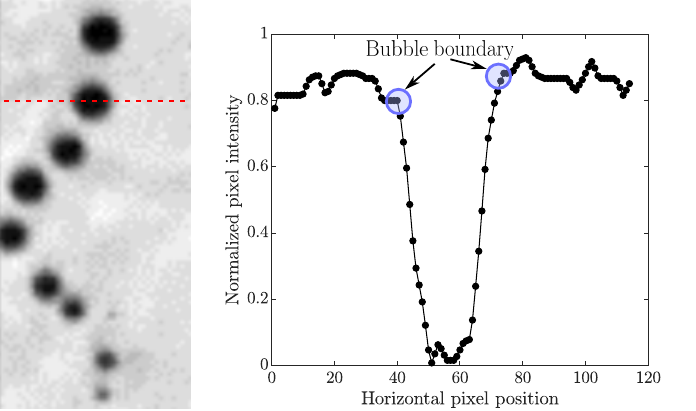}}
  \caption{Shadowgraphic image of bubble on the surface (left) and the intensity profile of the image at the cross-section drawn through the middle of one bubble (right). Dashed red line in the left image indicates the position of the cross-section.}
\label{fig:intensity}
\end{figure}

The uncertainty of the bubble radius measurement is associated with the limited spatial resolution of the optical setup, and are quantified by analysing the pixel intensity profiles of the recorded images. An example of a pixel intensity profile is shown in figure \ref{fig:intensity}, which shows the pixel intensity along the central cross section of a bubble. It can be seen that the transition from the dark interior of the bubble to the bright background happens within a span of about ten pixels. Therefore, the uncertainty of the radius is assumed to be $\pm 5$ pixels. The uncertainty of the bubble centroid is assumed to be small in comparison (less than one pixel), because the pixel intensity profile is symmetric about the bubble centre, which means that the position of the centroids will be the same no matter what threshold value is used to binarize the images. The nucleation site position is calculated as the initial position of newly nucleated bubbles; therefore, the uncertainty in the nucleation site position is also assumed to be relatively small. 

\subsection{Further justification of spherical bubble assumption and drag coefficient}
To justify the assumption that the bubble remains approximately spherical (section 3.2), equation \ref{bubble_eom} is used to calculate the Bond, Capillary and Weber number at departure for different mass fluxes and pressures. The velocity of the liquid surrounding the bubble is assumed to follow the law of the wall velocity profile, where the velocity is scaled by the shear (or friction) velocity $U_\tau=\sqrt{\tau_\textrm{w}/\rho_\textrm{l}}$. The wall shear stress is estimated from the McAdams shear stress correlation $\tau_\textrm{w}=0.023 Re_\textrm{ch}^{-0.2}G^{2}/ \rho_\textrm{l}$. The law of the wall velocity profile is calculated with the \citet{van_driest_turbulent_1956} eddy diffusivity model, yielding the turbulent kinematic viscosity $\nu_\textrm{t}=\nu_\textrm{l}f(y^{+})$. The resulting liquid velocity equation is then given by:
\begin{equation}
  U_\text{l}=U_{\tau}\int_{0}^{y^{+}}{\frac{dy^{+}}{1+\nu_\textrm{t}/\nu_\textrm{l}}},
    \label{U_l_integral}
\end{equation}
which can be simplified to a linear or logarithmic function depending on the magnitude of $y^{+}$.

\begin{figure}
  \centerline{\includegraphics[width=\textwidth]{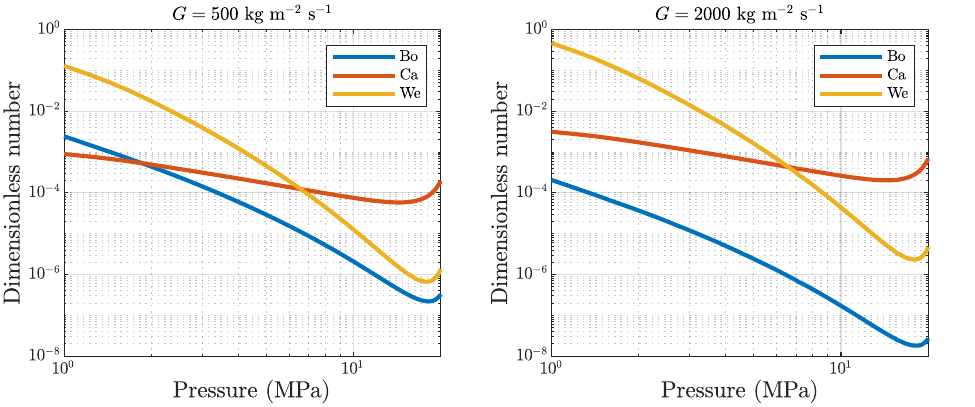}}
  \caption{Bond, Capillary and Weber number as a function of pressure at two different mass fluxes.}
\label{fig:BoCaWe}
\end{figure}

Results are shown in figure \ref{fig:BoCaWe}, where all dimensionless quantities are far less than unity at the mass fluxes and pressures used in the present study. Even at mass fluxes as high as 2000 kg m$^{-2}$ s$^{-1}$, the Capillary and Weber number are still much less than unity, indicating that surface tension forces should keep the bubble spherical in shape.
Deformation effects from turbulence and shear forces should also be negligible for these values of Bond, Capillary, and Weber number. \citet{Taylor1934} provided an analytical solution to predict the deformation of a bubble or droplet:
\begin{equation}
  \frac{R_\textrm{max}-R_\textrm{min}}{R_\textrm{max}+R_\textrm{min}}=\frac{19\mu_\textrm{v}+16\mu_\textrm{l}}{16\mu_\textrm{v}+16\mu_\textrm{l}}\frac{\tau_\textrm{w}D_\textrm{b}}{\sigma}
    \label{Taylor_deformation}
\end{equation}
and has been confirmed to agree well with recent direct numerical simulation results \citep{Soligo2020}. Note that equation \ref{Taylor_deformation} depends on a parameter resembling the standard Capillary number, which in our conditions is on the order of $10^{-3}$. Additionally, the work of \citet{Ni2024} suggests that our bubble shapes should hardly be affected by the turbulent environment, provided the turbulent Weber number, shear Weber number and Bond number are sufficiently small. In our conditions, we estimate the turbulent Weber number $We_\textrm{t}$ with the global energy dissipation rate ($\tau_\textrm{w}G/\rho_\textrm{l}^{2}D_\textrm{h}$) and find $We_\textrm{t}$ to be much less than unity, which should be small enough for turbulence-induced deformations to be negligible. We acknowledge that these estimations do not account for anisotropy in turbulence and variations in the energy dissipation rate as the bubble grows, and so turbulence induced deformations may need to be re-evaluated in future numerical studies.

\begin{figure}
  \centerline{\includegraphics[width=\textwidth]{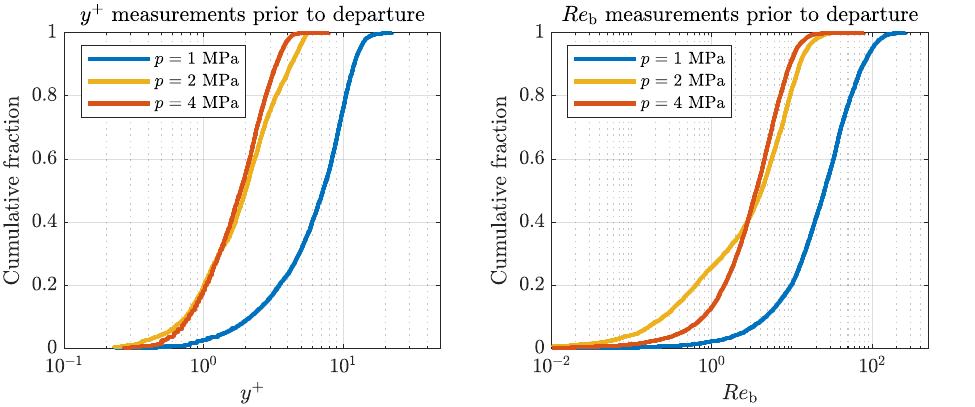}}
  \caption{Cumulative distribution function of the measured bubble diameter in wall units (left) and bubble Reynolds number (right) at different operating pressures.}
\label{fig:bubble_cdfs}
\end{figure}

\begin{figure}
  \centerline{\includegraphics[width=0.75\textwidth]{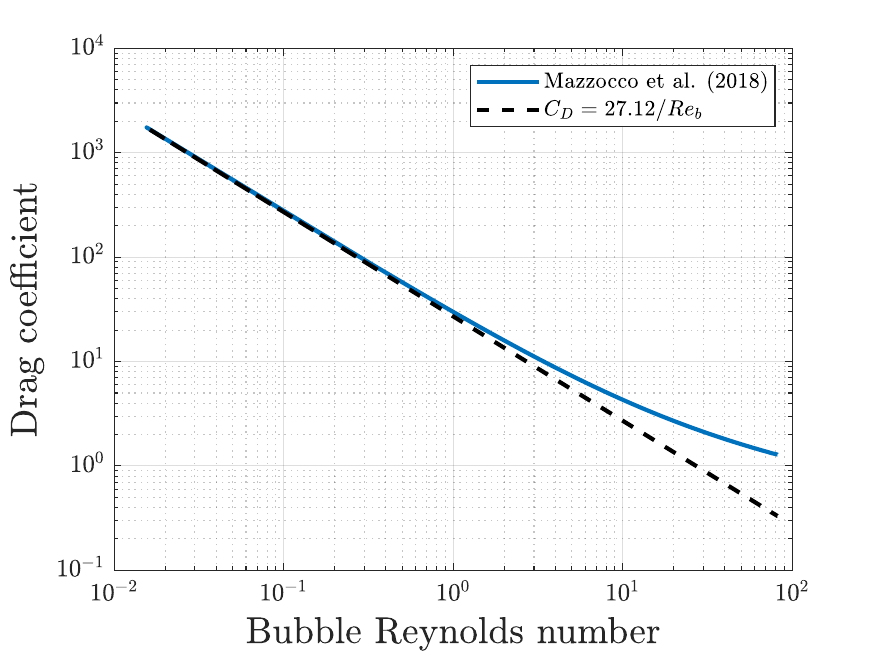}}
  \caption{Drag coefficient comparison.}
\label{fig:C_D_check}
\end{figure}

\begin{figure}
  \centerline{\includegraphics[width=\textwidth]{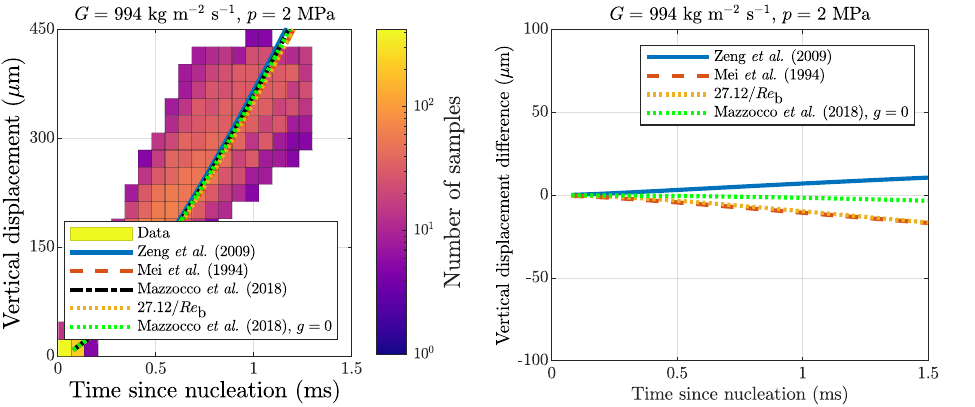}}
  \caption{Sensitivity of drag coefficient and buoyancy on bubble sliding predictions.}
\label{fig:sensitivity}
\end{figure}

The choice and sensitivity of drag coefficient depends on the bubble shape, Reynolds number, and distance from the wall. The bubble shape has been well established as spherical, but the drag coefficient we assume is only strictly valid if the bubble remains in the viscous sublayer and has a low Reynolds number. Cumulative distribution functions of our measured dimensionless bubble sizes $y^{+}$ and $Re_\textrm{b}$ are shown in figure \ref{fig:bubble_cdfs}. At high pressures, the vast majority of bubbles are within the viscous sublayer, and the bubble Reynolds number are typically on the order of 10 or less. Therefore, most bubbles are indeed inside a linear shear flow bounded by a wall and have a drag coefficient that is well approximated by equation \ref{C_D_simple}, as shown in figure \ref{fig:C_D_check}.
It can be seen that simplified drag model starts to deviate at a bubble Reynolds number of about 10, near the upper bound of measured $\Rey_\textrm{b}$ at high pressures. Ignoring nonlinear effects, the discrepancy in the drag coefficient would cause the bubble velocity in the simulations to match the liquid velocity more closely, because the $C_\text{D}$ is larger which would make the quantity $\Pi_1$ larger as well. Therefore, the analytical solution acts as a lower bound for the bubble velocity, while the liquid velocity is always a strict upper bound.
Finally, the sensitivity on the choice of drag coefficient was evaluated. An alternative drag coefficient correlation that seems suitable is the curve fit developed by \citet{zeng_forces_2009} in the case of significant surface contamination. This correlation and the drag coefficient for a clean spherical bubble in a uniform flow \citep{Mei1994} are used to predict the bubble trajectory for the sake of comparison. The results are shown in figure \ref{fig:sensitivity}, where the left graph shows different bubble trajectories are overlayed on each other, and the right graph shows the difference in displacement between the model in question and the model that utilizes the \citet{mazzocco_reassessed_2018} correlation (i.e., the force balance model used in the present study). Additionally, the case of zero buoyancy is also included to show the effect of buoyancy on sliding. It can be seen that nearly all model predictions lie on top of each other, with the largest difference ($\sim$10\%) between the models occurring beyond 1 ms, well beyond the time of departure. It is also clear that the choice of drag coefficient impacts the model predictions more (about one order of magnitude) than the inclusion of buoyancy, but not dramatically. 
\subsection{Evaluating different types of departure diameter schemes}
\begin{figure}
  \centerline{\includegraphics[width=0.75\textwidth]{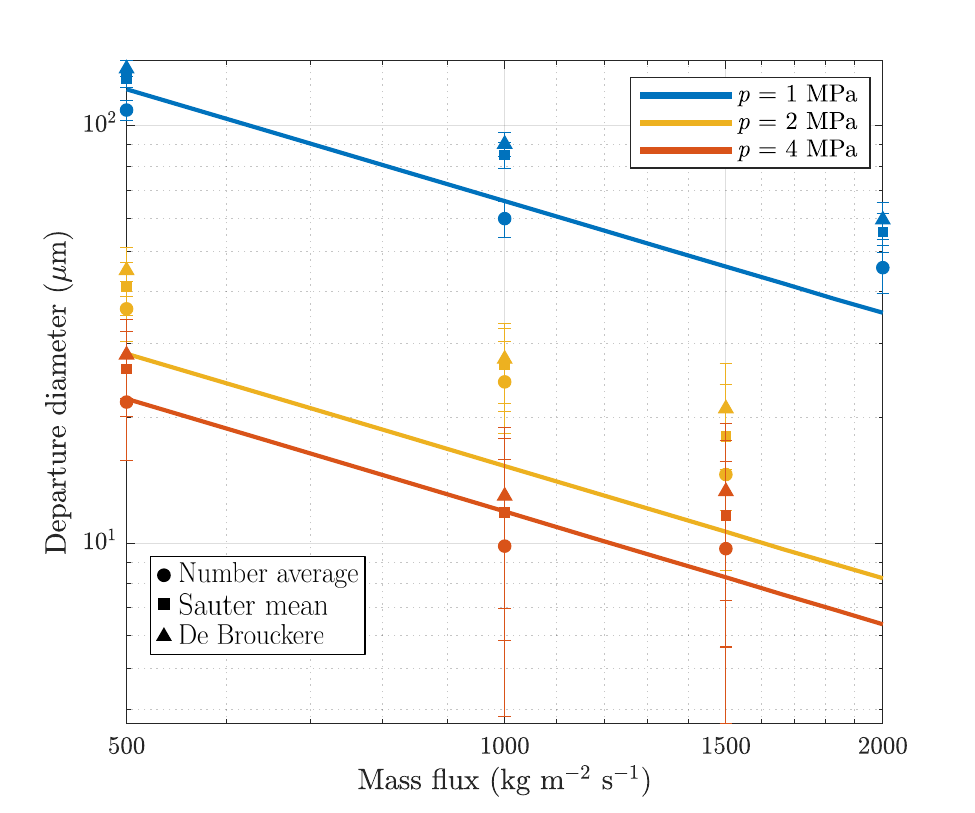}}
  \caption{Sensitivity of diameter averaging scheme on bubble departure diameter predictions.}
\label{fig:Dd_comp_all}
\end{figure}

Under each experimental operating condition, a distribution of measured departure diameters emerges. However, most modeling frameworks rely solely on a single representative departure diameter value. Therefore, in the present study we take the arithmetic average of the measured departure diameter data as the most representative value; however, number averaging is not the only approach. Other common averaging schemes to quantify an average particle size include the Sauter mean diameter and De Broukere mean diameter, given by:
\begin{equation}
    D_\text{d,32}=\frac{\sum_{i=1}^{n} D_i^{3}}{\sum_{i=1}^{n} D_i^{2}}
\end{equation}
and
\begin{equation}
    D_\text{d,43}=\frac{\sum_{i=1}^{n} D_i^{4}}{\sum_{i=1}^{n} D_i^{3}},
\end{equation}
respectively. A comparison between the three different departure diameter measurements is shown in figure \ref{fig:Dd_comp_all}. As shown, the discrepancy between the three averaging schemes is not substantial; therefore, for the sake of simplicity, we use the number average departure diameter in this study.
\clearpage
\bibliographystyle{jfm}
\bibliography{jfm}
\end{document}